\documentclass[sigconf]{acmart}

\usepackage{balance} 
\usepackage{times}
\usepackage{soul}
\usepackage{url}
\usepackage[utf8]{inputenc}
\usepackage{graphicx}
\usepackage{subfigure}
\usepackage{amsmath}
\usepackage{amsthm}
\usepackage{booktabs}
\usepackage{algorithm}
\usepackage[switch]{lineno}

\usepackage{bm}
\usepackage[noend]{algpseudocode}
\usepackage{xspace}
\usepackage{makecell}
\usepackage{multirow}
\usepackage{array}
\usepackage{color}
\usepackage{enumitem}
\usepackage{pifont}
\newcommand{\etal}{\emph{et al.}\xspace}
\newcommand{\eg}{\emph{e.g.,}\xspace}
\newcommand{\ie}{\emph{i.e.,}\xspace}

\newcommand{\baby}{GPFedRec\xspace}

\copyrightyear{2024}
\acmYear{2024}
\setcopyright{acmlicensed}
\acmConference[KDD '24] {Proceedings of the 30th ACM SIGKDD Conference on Knowledge Discovery and Data Mining }{August 25--29, 2024}{Barcelona, Spain.}
\acmBooktitle{Proceedings of the 30th ACM SIGKDD Conference on Knowledge Discovery and Data Mining (KDD '24), August 25--29, 2024, Barcelona, Spain}
\acmISBN{979-8-4007-0490-1/24/08}
\acmDOI{10.1145/3637528.3671702}

\begin{document}

%%
%% The "title" command has an optional parameter,
%% allowing the author to define a "short title" to be used in page headers.
\title{\baby: Graph-Guided Personalization for Federated Recommendation}

%%
%% The "author" command and its associated commands are used to define
%% the authors and their affiliations.
%% Of note is the shared affiliation of the first two authors, and the
%% "authornote" and "authornotemark" commands
%% used to denote shared contribution to the research.
\author{Chunxu Zhang}
\affiliation{%
  \institution{College of Computer Science and Technology, Jilin University}
  \institution{Key Laboratory of Symbolic Computation and Knowledge Engineering of Ministry of Education, Jilin University}
  \country{Changchun, China}
}
\email{cxzhang19@mails.jlu.edu.cn}

\author{Guodong Long}
\affiliation{%
  \institution{Australian Artificial Intelligence Institute, FEIT, University of Technology Sydney}
  \country{Sydney, Australia}
}
\email{guodong.long@uts.edu.au}

\author{Tianyi Zhou}
\affiliation{%
  \institution{Computer Science and UMIACS, University of Maryland}
  \country{Maryland, USA}
}
\email{zhou@umiacs.umd.edu}

\author{Zijian Zhang}
\affiliation{%
  \institution{College of Computer Science and Technology, Jilin University}
  \institution{City University of Hong Kong}
  \country{China}
}
\email{zhangzj2114@mails.jlu.edu.cn}

\author{Peng Yan}
\affiliation{%
  \institution{Australian Artificial Intelligence Institute, FEIT, University of Technology Sydney}
  \country{Sydney, Australia}
}
\email{yanpeng9008@hotmail.com}

\author{Bo Yang}
\affiliation{%
  \institution{College of Computer Science and Technology, Jilin University}
  \institution{Key Laboratory of Symbolic Computation and Knowledge Engineering of Ministry of Education, Jilin University}
  \country{Changchun, China}
}
\email{ybo@jlu.edu.cn}
\authornote{Corresponding author.}

%%
%% By default, the full list of authors will be used in the page
%% headers. Often, this list is too long, and will overlap
%% other information printed in the page headers. This command allows
%% the author to define a more concise list
%% of authors' names for this purpose.
\renewcommand{\shortauthors}{Chunxu Zhang et al.}

%%
%% The abstract is a short summary of the work to be presented in the
%% article.
\begin{abstract}
  The federated recommendation system is an emerging AI service architecture that provides recommendation services in a privacy-preserving manner. Using user-relation graphs to enhance federated recommendations is a promising topic. However, it is still an open challenge to construct the user-relation graph while preserving data locality-based privacy protection in federated settings. Inspired by a simple motivation, similar users share a similar vision (embeddings) to the same item set, this paper proposes a novel Graph-guided Personalization for Federated Recommendation (\baby). The proposed method constructs a user-relation graph from user-specific personalized item embeddings at the server without accessing the users' interaction records.  The personalized item embedding is locally fine-tuned on each device, and then a user-relation graph will be constructed by measuring the similarity among client-specific item embeddings. Without accessing users' historical interactions, we embody the data locality-based privacy protection of vanilla federated learning. Furthermore, a graph-guided aggregation mechanism is designed to leverage the user-relation graph and federated optimization framework simultaneously. Extensive experiments on five benchmark datasets demonstrate \baby's superior performance. The in-depth study validates that \baby can generally improve existing federated recommendation methods as a plugin while keeping user privacy safe. Code is available to ease reproducibility\footnote{\url{https://github.com/Zhangcx19/GPFedRec}}.
\end{abstract}

%%
%% The code below is generated by the tool at http://dl.acm.org/ccs.cfm.
%% Please copy and paste the code instead of the example below.
%%
\begin{CCSXML}
<ccs2012>
   <concept>
       <concept_id>10002951</concept_id>
       <concept_desc>Information systems</concept_desc>
       <concept_significance>500</concept_significance>
       </concept>
   <concept>
       <concept_id>10002951.10003317.10003331.10003271</concept_id>
       <concept_desc>Information systems~Personalization</concept_desc>
       <concept_significance>500</concept_significance>
       </concept>
 </ccs2012>
\end{CCSXML}

\ccsdesc[500]{Information systems}
\ccsdesc[500]{Information systems~User modeling}

%%
%% Keywords. The author(s) should pick words that accurately describe
%% the work being presented. Separate the keywords with commas.
\keywords{Federated Learning; Recommendation Systems; User Graph}
%% A "teaser" image appears between the author and affiliation
%% information and the body of the document, and typically spans the
%% page.

%%
%% This command processes the author and affiliation and title
%% information and builds the first part of the formatted document.
\maketitle

\section{Introduction}\label{introduction}
In the era of the information explosion, people are overwhelmed by the data with boosting volume. To address this challenge, recommendation systems have become essential in discovering users' interests and filtering out their unconcerned content.  However, existing recommendation models rely on centralized user data storage, which risks privacy violations and has attracted increasing social concerns, \eg General Data Protection Regulation (GDPR)~\cite{voigt2017eu}. As an emerging service architecture, the federated recommendation system has been proposed to provide recommendations while preserving user privacy~\cite{chai2020secure,perifanis2022federated,qu2023semi,zhang2023dual,yin2024device}. It usually trains the recommendation model on the local user device (\ie client), and a server orchestrates the training process by synchronizing the shared model parameters. Privacy protection can be achieved since the user data is preserved on each client locally and cannot be accessed by others.

Existing federated recommendation research generally treats users as individuals to train the global model, while overlooking the correlations between them. In recommendation scenarios, users usually have diverse connections. For instance, users who have purchased the same items exhibit common interests and may also prefer the other same product \cite{schafer2007collaborative}. These correlations can be effectively described with a graph structure~\cite{guo2020survey,wu2022graph,gao2023survey,pei2024memory}. Using it in the recommendation systems can enrich the user (item) representation learning and promote user preference modeling, which has become a popular paradigm and achieved outstanding performance in the centralized setting~\cite{he2020lightgcn,wang2019kgat,wu2021self}. Hence, developing a user relationship graph-enhanced federated recommendation system holds the potential to provide better privacy-preserving recommendation services.

\begin{figure}[!t]
\setlength{\belowcaptionskip}{-8pt}
\includegraphics[scale=0.26]{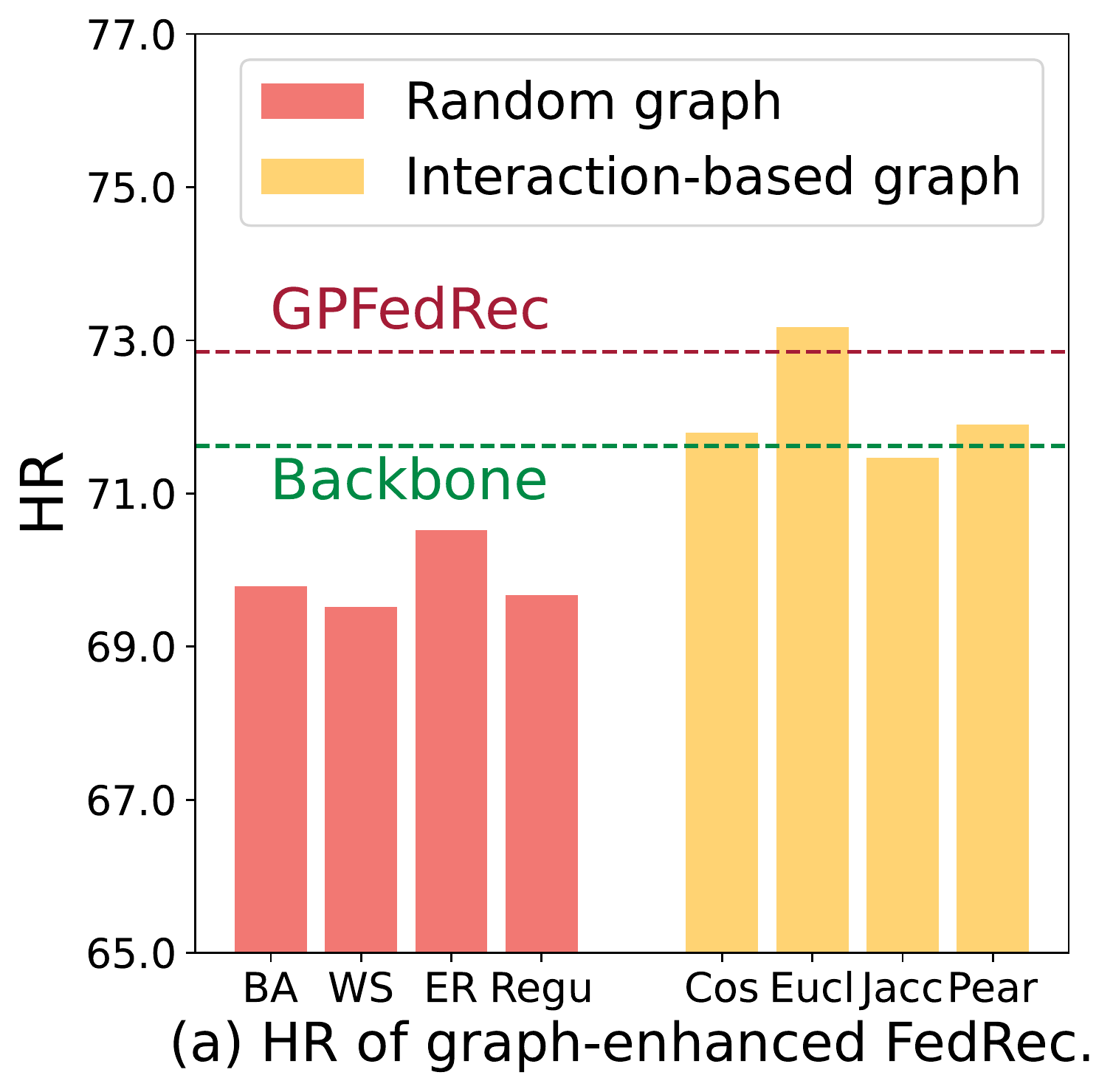}
% \hspace{0.2in}
\includegraphics[scale=0.26]{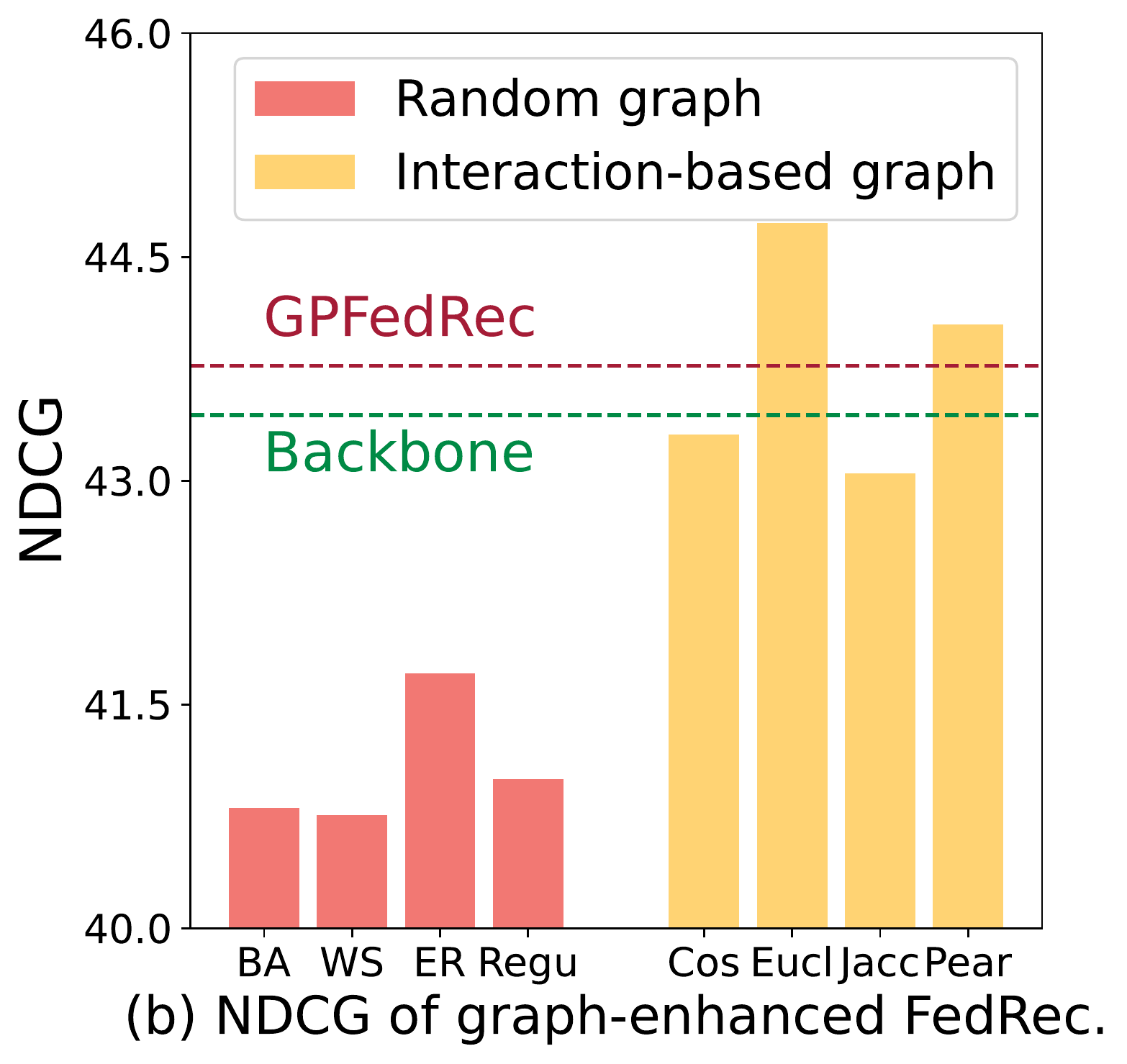}
\caption{Performance comparison of diverse user relationship graphs-enhanced federated recommendation model on the MovieLens-100K dataset. {Backbone} denotes the current state-of-the-art federated recommendation model. 
% More details can be found in Section Experiment.
}
\label{diff_graphs}
\end{figure}

To this end, we first study how to integrate user-relation graph into federated recommendation system. Specifically, we adopt two straight and widely used methods to construct the graph, \ie randomly generated graph and built based on user historical interactions, details can be referred to section~\ref{exp_diff_graph}. Then, we conduct a preliminary experiment to analyze the contribution of the user relationship graph to the system performance. Particularly, we equip the current state-of-the-art federated recommendation model PFedRec~\cite{zhang2023dual} with the former two kinds of graph, and compare the performance on the MovieLens-100K dataset. As shown in Figure~\ref{diff_graphs}, random graphs hurt the performance, while the informative graphs built with interactions can improve the performance. However, the user interaction data are private and cannot be accessed to build the graph. Hence, the challenge of developing a user relationship graph-enhanced federated recommendation model lies in \textbf{building an informative graph without increasing the risk of user privacy leakage.}

In this paper, we present a novel \textbf{G}raph-guided \textbf{P}ersonalization framework for \textbf{Fed}erated \textbf{Rec}ommendation (\textbf{\baby}), which is the first user correlation-enhanced general framework for modeling personalized federated recommendation system. Within the federated optimization paradigm, the server can obtain model parameters learned by individual clients based on their historical interaction data. These parameters encompass user characteristics and can ensure user privacy security, making them viable information for constructing the user-relation graph. In order to enhance the user modeling with correlated users, we propose to construct the user relationship graph on the server with locally updated item embeddings. This mechanism effectively identifies users' relationships while relieving user's private data from exposure. Furthermore, we design a novel graph-guided aggregation mechanism to exploit the user correlations in global model parameter aggregation. Thus, the server can learn user-specific instead of indiscriminate item embeddings, which are then assigned to clients to promote user personalization capture.

We evaluate \baby's performance on five recommendation benchmark datasets and compare it with advancing baselines. Experiments demonstrate that \baby consistently achieves state-of-the-art performance. Then, we conduct ablation studies to analyze the effect of user-relation graph on our model. As illustrated in Figure~\ref{diff_graphs}, our proposed graph construction method can achieve comparable performance with interaction-based graph and outperform the random graph significantly. To further verify the effectiveness and compatibility of our method, we also enhance other FedRec methods with our graph-guided aggregation mechanism. The results show that our graph-guided aggregation mechanism can generally improve federated recommendation methods as a swift plugin. Besides, we incorporate the differential privacy technique to further enhance privacy protection and empirical results show that \baby achieves a steady performance under the privacy-preserving scenario, which supports the practical feasibility. In a nutshell, our \textbf{main contributions} are summarized as follows,

\begin{itemize}[leftmargin=*]
    \item We present a novel approach to identify correlations among users in the federated recommendation setting, which constructs the user relationship graph with the shared item embeddings without privacy exposure.

    \item We introduce a graph-guided aggregation mechanism that enables the learning of user-specific item embeddings, thus promoting user personalization modeling. The overall algorithm can be formulated into a unified federated optimization framework \baby.
    
    \item The proposed method achieves state-of-the-art performance on five recommendation benchmark datasets, and extensive analyses verify its efficacy and privacy-preserving capability.
    
    \item Our simple yet effective graph-guided aggregation mechanism owns promising compatibility, which could generally enhance the existing federated recommendation methods as a swift plugin. 
\end{itemize}

\section{Related Work}\label{related work}
\subsection{Graph Learning-based Recommendation System}
Graph learning-based recommendation systems can learn enhanced user (item) embedding by explicitly exploiting high-order neighbor information in the graph structure, which has been a burgeoning paradigm. Integrating the user-item interaction graph into the collaborative filtering framework is a straightforward strategy. For example, He \etal propose LightGCN~\cite{he2020lightgcn}, which applies the graph convolution network to the user-item interaction graph to enrich representation learning for user preference prediction. By considering the adjacency between items, the interacted item sequences can be organized as a graph. Correspondingly, the sequential recommendation can be achieved by capturing the transition pattern from the sequence graph~\cite{pan2020star,latifi2022streaming}. With the emergence of social networks, the social recommendation system is developed to enhance user modeling by means of local neighborhoods. The basic assumption is that users with social relationships should have similar representations. Generally, existing works either take the user-item interaction graph and social network as two graphs to learn user representation separately~\cite{wu2019neural} or integrate both graphs into the unified graph to learn enhanced user representation~\cite{wu2020diffnet++}. Besides, there are also some works developing the recommendation models based on the knowledge graph~\cite{wang2021learning,yang2022knowledge,zou2022multi}, which introduces side information into the graph, \eg item features. Existing methods collect user data centrally, which violates user privacy protection. This paper proposes a recommendation model based on the federated learning framework, combined with privacy protection technology to protect users' private data from exposure.

\subsection{Federated Recommendation Systems}
Federated Recommendation (FedRec)~\cite{li2023federated,yuan2023interaction,zhang2023comprehensive,zhang2024federated} is an emerging direction that aims to provide recommendation services without collecting private user data under the federated learning framework~\cite{miao2023task,zhong2023personalized}. Various benchmark recommendation methods have been adapted to the federated learning framework, such as matrix factorization-based FCF~\cite{ammad2019federated}, FedMF~\cite{chai2020secure}, MetaMF~\cite{lin2020meta} and FedRecon~\cite{singhal2021federated} and neural collaborative filtering-based FedNCF~\cite{perifanis2022federated}. Zhang \etal \cite{zhang2023dual} presents a personalized FedRec framework named PFedRec, which removes the user embedding and learns the personalized score function to capture user preference. However, these methods neglect the correlations among users, which are commonly used information in centralized recommendation settings. To bridge this gap,  Liu \etal~\cite{liu2022federated} propose to enhance the local subgraph by introducing the user social relationship information. However, the social network is not always accessible in practical scenarios. Wu \etal \cite{wu2022federated} present the FedPerGNN, which organizes the local user-item interactions as a graph and deploys a graph neural network on each client to capture the user-item correlations. Besides, FedPerGNN employs a third-party server to find the high-order neighbors so as to provide more beneficial information for local model training. However, aligning the historical interactions with the third-party server results in high computational overhead and increases the risk of user privacy information exposure. Furthermore, existing FedRec models generally learn shared model parameters for all users, which neglects the diverse user preferences. In this paper, we design a graph-guided aggregation mechanism to capture user preference correlations, which promotes the personalized FedRec system.

\begin{figure*}[!t]
    \centering
    \includegraphics[width=0.9\textwidth,height=0.45\textwidth]{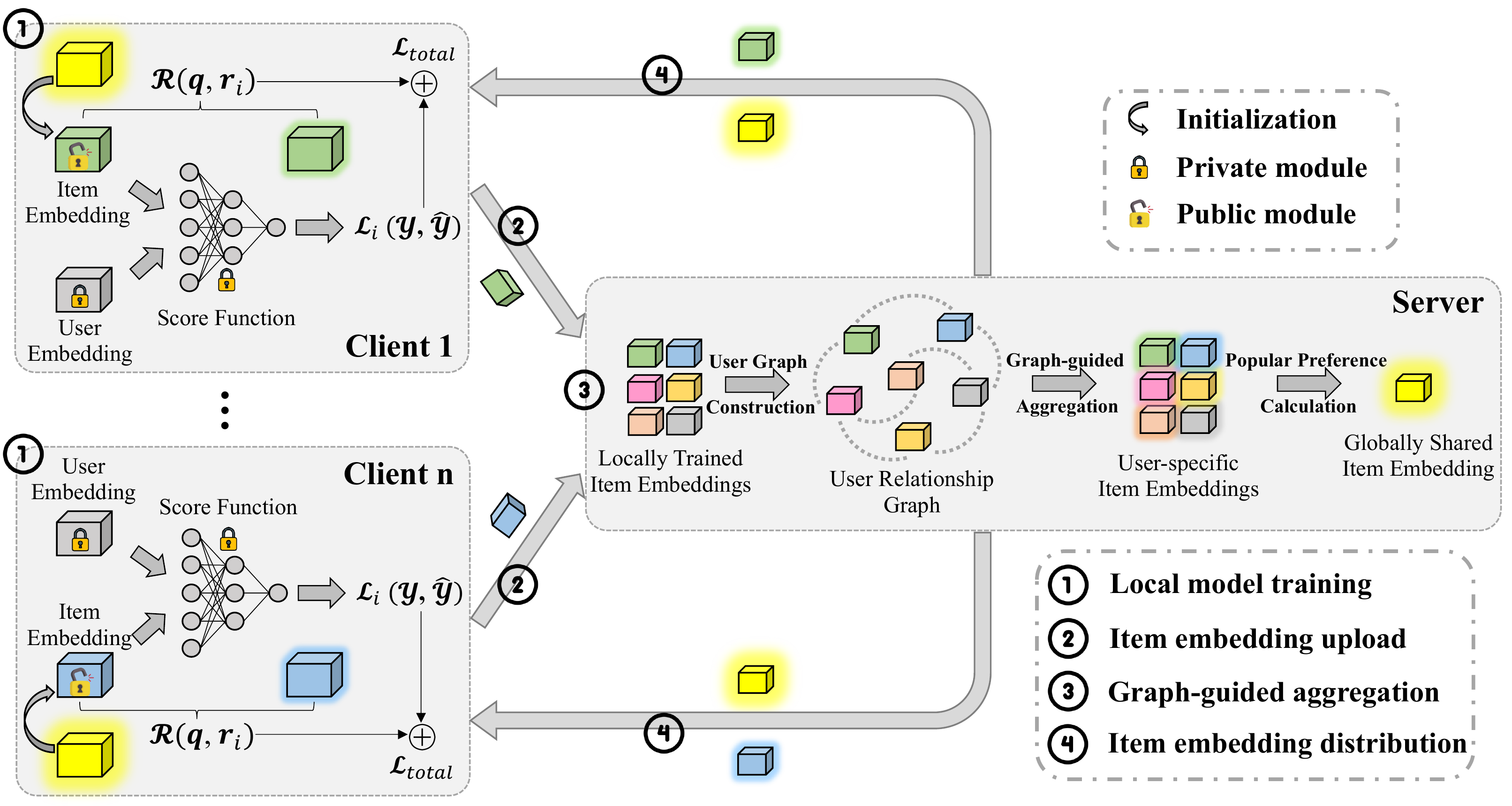}
    \caption{The framework of \baby. There are four steps in each communication round: \ding{172} For the local recommendation model trained on each client, it initializes the item embedding with the globally shared item embedding. Then it takes the user-specific item embedding as a regularizer $\mathcal{R}(q,r_i)$ together with the loss of the recommendation task $\mathcal{L}_i(\mathcal{Y},\hat{\mathcal{Y}})$ as the optimization objective $\mathcal{L}_{total}$. \ding{173} The client uploads the locally updated item embedding $q$ to the server. \ding{174} For the server, it first constructs a user relationship graph based on the received item embeddings. Then, it performs the graph-guided aggregation to achieve user-specific item embeddings $\{r_i\}_{i=1}^n$ and meanwhile calculates the globally shared item embedding depicting the popular preference. \ding{175} The server distributes both the globally shared and user-specific item embeddings to the clients for the next round of optimization.}
    \label{fig:framework}
\end{figure*}

\section{Preliminary}\label{preliminary}
\begin{table}[!t]
\renewcommand\arraystretch{1.}
\setlength\tabcolsep{0.8pt}
\setlength{\abovecaptionskip}{-15pt}
\centering
% \normalsize
\begin{tabular}{p{35pt}p{200pt}<{\centering}}
\hline
\pmb{Notation} & \pmb{Descriptions} \\
\hline
$\mathcal{U}$ & The user set \\
$\mathcal{I}$ & The item set \\
$\mathcal{Y}_{um}$ & The rating of user $u$ on item $m$ \\
$\hat{ \mathcal{Y}}_{um}$ & The prediction of score function \\
$\mathcal{G}(\mathcal{U}, \mathcal{E})$ & The user relationship graph \\
$\mathcal{A}$ & The adjacency matrix of user relationship graph \\
$\mathcal{S}$ & The user similarity matrix \\
$\mathcal{M}_\theta$ & The recommendation model parameterized with $\theta$ \\
$N$ & The number of clients (users) \\
$p_i$ & The user embedding module parameter of $i$-th client \\
$q_i$ & The item embedding module parameter of $i$-th client \\
$o_i$ & The score function module parameter of $i$-th client \\
$r_i$ & The user-specific item embedding of $i$-th client \\
$q_{global}$ & The globally shared item embedding \\
\hline
\end{tabular}
\caption{Notation table.}
\label{notation}
\end{table}

\noindent\textbf{Federated Recommendation.}
Let $\mathcal{U}$ and $\mathcal{I}$ represent the user and item sets, respectively. Let $\mathcal{Y}_{um}$ be the user-item interaction data, indexed by user $u$ and item $m$.
Other notations could be referred to Table \ref{notation}.
For a recommendation model $\mathcal{F}$ parameterized by $\theta$, it makes prediction as $\hat{ \mathcal{Y}}_{um}=\mathcal{F}(u,m|\theta)$.
Denote user relationship graph with $\mathcal{G}(\mathcal{U}, \mathcal{E})$, where $\mathcal{U}$ and $\mathcal{E}$ are the sets of users and edges, respectively. Denote the adjacency matrix of $\mathcal{G}$ by $\mathcal{A} \in \{0, 1\}^{N\times N}$ where $N$ is the number of user in $\mathcal{U}$. $\mathcal{A}_{uv}=1$ indicates an edge between users $u, v \in \mathcal{U}$, otherwise $\mathcal{A}_{uv}=0$.

For a model $\mathcal{F}$ parameterized by $\theta$, federated recommendation aims to predict user $u$'s preference on item $m$ as $\hat{ \mathcal{Y}}_{um}=\mathcal{F}(u,m|\theta^*)$, and the optimal model parameter $\theta^*={\rm argmin}_\theta \sum_{i=1}^{N}\omega_i \mathcal{L}_i(\theta)$ is learned by minimizing the accumulated loss of all local models $\mathcal{L}_i(\theta)$ with client weight $\omega_i$.

\section{Methodology}\label{method}
In this section, we present the \textbf{G}raph-guided \textbf{P}ersonalization framework for \textbf{Fed}erated \textbf{Rec}ommendation (\textbf{\baby}). As shown in Figure \ref{fig:framework},
we address the intrinsic relationship between users with a graph structure. In each training round, the server first gathers locally trained item embeddings from clients. Then, the server updates them with a graph-guided aggregation mechanism, which achieves user-specific item embeddings. Meanwhile, a shared item embedding is calculated to depict the popular preference. Finally, both user-specific and shared item embeddings are distributed to clients for personalized local model learning.

In the following part, we introduce the proposed \baby in detail. We first formulate the overall objective function under the federated learning framework. Then, we illustrate the local recommendation model loss function of each client. In addition, we detail the learning process and summarize the overall optimization workflow into an algorithm. Furthermore, we analyze the privacy-preserving capability of our method and the further enhancement by integrating privacy protection techniques. Finally, we discuss the efficiency and scalability of \baby and the potential extension of it on more general recommendation scenarios.

\subsection{Federated Optimization Objective}\label{optimization objective}
We consider each user as a client in the federated learning framework. The recommendation task can be described as a personalized federated learning problem, which aims to provide personalized service for each user. We employ a neural recommendation model $\mathcal{M}_\theta$, which contains three components, including a user embedding module parameterized by $p$, an item embedding module parameterized by $q$ and a score function module parameterized by $o$ that predicts user's rating based on user and item embeddings.

Particularly, we assign item embedding as a shared role, which is responsible for transferring common knowledge among users. Both the user embedding and the score function are maintained locally to capture user personalization. We formulate the proposed \baby as the below optimization objective,
\begin{equation} \label{eq:global-loss}
  \begin{aligned} 
    &\min_{\{\theta_1,...,\theta_N\}} & \sum_{i=1}^N \mathcal{L}_i(\theta_i) + \lambda \mathcal{R}(q_i, r_i) \\
    % &s.t. & q^* \in \arg\min_{q}  \sum_{i=1}^{N}\omega_i \mathcal{L}_i(q) \\
    % &s.t. &r_i := \sum_{j \in \mathcal{N}_{(i)}} q_j
  \end{aligned}
\end{equation}
where $\theta_i=\{p_i,q_i,o_i\}$ is the recommendation model parameter of $i$-th client. $r_i$ is the user-specific item embedding learned on the server. $\mathcal{R}(\cdot,\cdot)$ is a regularization term to constrain the local item embedding to be similar to user-specific item embedding, and $\lambda$ is the regularization coefficient. 

\subsection{Recommendation Model Loss Function}
To pursue the generality, we discuss the typical scenario where recommendation only relies on the implicit user-item interaction data, \ie $\mathcal{Y}_{um}=1$ if user $u$ has interacted with item $m$; otherwise, $\mathcal{Y}_{um}=0$. No auxiliary user (item) raw features are available. Due to the binary value of implicit feedback, we define the loss function for the $i$-th client as the \textit{binary cross-entropy loss},
\begin{equation} \label{eq:local-loss}
    \mathcal{L}_i(\theta_i;\mathcal{Y}_{um},\hat{\mathcal{Y}}_{um}) = - \sum_{(u,m) \in D_i} \log \hat{\mathcal{Y}}_{um} - \sum_{(u,m') \in D_i^-} \log (1 - \hat{\mathcal{Y}}_{um'})
\end{equation}
where $D_i$ and $D_i^-$ denote the interacted positive item set and sampled negative item set of $i$-th client, respectively. The $\hat{\mathcal{Y}}_{um}$ is the model prediction. For efficient $D_i^-$ construction, we sample negative instances from the user's unobserved interaction collection according to the negative sampling ratio. 

\subsection{Optimization}
To solve the optimization objective in Eq. (\ref{eq:global-loss}), we conduct below two alternate steps. \textbf{\textit{First}}, the server learns the user-specific item embeddings $r_i$ based on the graph-guided aggregation mechanism, and meanwhile achieves a global item embedding $q_{global}$ depicting popular preferences. \textbf{\textit{Second}}, we update $\theta_i$ initialized with $q_{global}$ by solving the local loss function in Eq. (\ref{eq:local-loss}) with a regularization term $\mathcal{R}(q_i, r_i)$: distance between local item embedding and user-specific item embedding. Details of the two steps are introduced next.

\subsubsection{\textbf{Server update with graph-guided aggregation.}}
The server receives item embeddings from clients, which are learned with the personal interaction data, and the relationship with other clients is missing. Besides, the vanilla federated learning framework, \eg FedAvg~\cite{mcmahan2017communication}, treats each client equally and learns a unified item embedding with average aggregation. However, we argue that the user generally shares similar preferences with a user group, and taking the average common preference from all users will hinder the user personalization modeling.

To capture the correlations among users and achieve user-specific preference capture, we propose to build a user relationship graph $\mathcal{G}$ on the server. Particularly, we identify the relevance between users by calculating the similarities of locally updated item embeddings. The insight behind this is that the users with common preferences share similar views of the items. Besides, it can be safely shared without disclosing private user information. 
% Meanwhile, users who connect with more users indicate that they hold popular preferences, which is generally beneficial information for recommendation systems.

Specifically, we employ the \textit{cosine similarity} as the similarity metric between item embeddings, and the similarity between client $i$ and $j$ can be formulated as,
\begin{equation}\label{cos}
    \mathcal{S}_{ij} = \frac{q_i \cdot q_j}{||q_i||||q_j||}
\end{equation}
where $q_i$ and $q_j$ are the item embeddings of the two clients.

Given the relationship indicator $\mathcal{S}_{ij}$, we select users with high similarity as the neighbors. While specifying the neighborhood size is difficult and tends to introduce redundant information. To overcome this issue, we devise a more flexible neighborhood selection strategy. We establish an adaptive threshold to decide neighbors, \ie users whose similarity is greater than the threshold are reserved as neighbors. Particularly, we take the mean value $\mathcal{\overline{S}}$ of the similarity matrix $\mathcal{S}$ as a reference,
\begin{equation}\label{graph}
    \mathcal{A}_{ij}=\begin{cases}
    1\;& \mathcal{S}_{ij}>\gamma \mathcal{\overline{S}} \\
    0\;& otherwise\\
\end{cases}
\end{equation}
where $\gamma$ is a scaling factor used to set the similarity threshold. During federated optimization, the item embeddings received from clients are updated consistently, and hence, the user relationship graph will be changed adaptively.

Based on the graph, we design a graph-guided aggregation mechanism to update the item embeddings so that each client can obtain the user-specific item embedding with the help of neighbors with similar preferences. Specifically, we employ a lightweight Graph Convolution Network (GCN) \cite{he2020lightgcn} to update the $i$-th client item embedding by aggregating its neighbors and obtain $r_i$, with the following convolution operation form,
\begin{equation}\label{gcn}
    R = \mathcal{A}^lQ
\end{equation}
where $Q$ is the initial item embedding matrix whose $i$-th row represents the item embedding received from user $i$ and $R$ is the aggregated item embedding matrix whose $i$-th row is $r_i$. Besides, $l$ indicates the number of convolution layers. For simplicity, in this paper, we use $l=1$ convolution layer. It is mentioned that the $\mathcal{A}$ can be replaced with other reasonable forms, \eg Laplace matrix. In this paper, we take the vanilla adjacent matrix. 

Under the graph-guided aggregation mechanism, users with more neighbors will participate in more $r_i$ calculations. To capture the popular preference, we employ a simple average on all $r_i$ for achieving shared item embedding, where the users with more neighbors will hold higher weights. We formulate the calculation in the following,
\begin{equation}\label{dq}
    q_{global} = \mathcal{D}Q
\end{equation}
where $\mathcal{D}$ is the degree matrix of $\mathcal{A}$ when $l=1$. Compared to indiscriminate aggregation of existing methods, our solution pays more attention to users with popular preferences, which achieves better performance displayed in empirical verification in the experiment.

\subsubsection{\textbf{Client update with regularization.}}
In each training round, the client receives two forms of item embedding from the server, including the shared $q_{global}$ depicting popular preference and the user-specific $r_i$ depicting personalized preference. We incorporate both forms of preference into local model training. Particularly, we first initialize the item embedding $q_i$ with global shared item embedding $q_{global}$ and both user embedding $p_i$ and score function $o_i$ are inherited from the trained model in the last round. Then, we train the model by regularizing $q_i$ close to the personalized item embedding $r_i$, which can be formulated as follows,
\begin{equation} \label{eq:local-loss-total}
    \mathcal{L}_{total} = \mathcal{L}_i(\theta_i;\mathcal{Y}_{um},\hat{\mathcal{Y}}_{um}) + \lambda \mathcal{R}(q_i, r_i)
\end{equation}
where $\lambda$ is the coefficient of the regularization term. We minimize the distance between $q_i$ and $r_i$ with the \textit{mean square error} as the loss function, \ie $\mathcal{R}(\cdot,\cdot)=MSE(\cdot,\cdot)$.

We update the $\theta_i$ with stochastic gradient descent algorithm, and $t$-th update step can be formulated as follows,
\begin{equation}\label{local update}
    \theta_i^t = \theta_i^t - \eta \partial_{\theta_i^t} \mathcal{L}_{total}
\end{equation}
where $\eta$ is the learning rate and $\partial_{\theta_i^t} \mathcal{L}_{total}$ is the gradient of model parameters with respect to loss.

\begin{algorithm}[!t]
% \setstretch{1.15}
\caption{Graph-guided Personalization for Federated Recommendation - Optimization Procedure}
% {\bf ServerExecute:}
\begin{algorithmic}[1]
\State Initialize $\lambda$, $\eta$, $\gamma$, $\{(p_i^{(1)}, q_i^{(1)}, o_i^{(1)})\}_{i=1}^N$
\State Initialize $\{r_i^{(1)}\}_{i=1}^N \leftarrow \{q_i^{(1)}\}_{i=1}^N$ 
\For{each round $t=1,2,...,T$}
\State \underline{\textit{Server update with graph-guided aggregation}}:
% \State $\{r_i^{(t+1)}\}_{i=1}^N \leftarrow \{q_i^{(t+1)}\}_{i=1}^N$
\State Calculate the similarities of locally updated item embeddings with Eq. (4)
\State Build user relationship graph $\mathcal{G(A)}^{(t)}$ with Eq. (5)
\State Learn user-specific item embeddings $\{r_i^{(t+1)}\}_{i=1}^N$ with Eq. (6)
\State Learn globally shared item embedding $q_{global}$ with Eq. (7)
\State \underline{\textit{Client update with regularization}}:
\For{each client $i=1,2,...,N$ in parallel}
% \State Initialize $q_i^{(t)}$ with $q_{global}^{(t)}$
% \State initialize $p_i^{(t)}$ and $o_i{(t)}$ with the 
\For{each epoch $e$ from 1 to $E$}
\State Update $(p_i^{(t)},q_i^{(t)},o_i^{(t)})$ with Eq. (9)
\EndFor
\State \textbf{end for}
\State $(p_i^{(t+1)},q_i^{(t+1)},o_i^{(t+1)}) \leftarrow (p_i^{(t)},q_i^{(t)},o_i^{(t)})$
\EndFor
\State \textbf{end for}
\EndFor
\State \textbf{end for}
\end{algorithmic}
\label{algorithm}
\end{algorithm}
\subsection{Algorithm}
\subsubsection{Overall optimization.}
The optimization objective can be solved iteratively through multiple communication rounds between the server and clients. In the beginning, we initialize the recommendation model $\mathcal{M}_\theta$ for all clients. For each communication round, the server updates the item embedding with the graph-guided aggregation mechanism, and distributes both the user-specific item embedding $r_i$ and globally shared item embedding $g_{global}$ to clients for the local update. Then, the client trains the local recommendation model with the personal interactions and uploads the updated item embedding $q_i$ to the server for the subsequent communication round. The overall optimization procedure is organized into Algorithm~\ref{algorithm}.

\subsubsection{Efficient item embedding storage on client.}
In the practical scenario, there will be a large number of items in the recommendation system, which brings potential item embedding storage and communication overhead challenges for client devices with constrained resources. To handle this issue, we advocate that each client can only preserve the interacted items and randomly sampled items, which are far less than the complete items, resulting in efficient item embedding storage on the client. To alleviate the high storage requirements during inference, the server can first filter the items that users may interested in (e.g., calculate the item similarities between updated items and other candidate items and select the candidate items with high similarities), and the clients only need to perform ranking on the item subset instead of the full item set.

\subsection{Privacy Protection Enhanced \baby}
Under the federated learning framework, our method inherits the privacy-preserving merit that each user preserves private data locally, which could significantly reduce the risk of privacy leakage. In terms of further handling the potential privacy violation when uploading item embedding to the server, we propose to integrate the local differential privacy strategy~\cite{choi2018guaranteeing} into our method. Particularly, we incorporate a zero-mean Laplacian noise to the item embedding before it is uploaded to the server,
\begin{equation}
    q_i = q_i + Laplacian(0,\delta)
\end{equation}
where $\delta$ is the noise intensity. Hence, one cannot easily obtain the updated items by monitoring item embeddings, and the privacy protection ability is better as $\delta$ increases.

\subsection{Discussions}\label{discussions}
\subsubsection{Efficiency and scalability about \baby.}
In the practical application, there are usually many clients in recommendation systems, which challenges the efficiency and scalability of our graph-guided aggregation on the server. To address the above challenges, we discuss the feasible solutions from user relationship graph construction and user-specific item embedding learning, respectively. The goal of user relationship graph construction is to discover the correlations among users. Generally, the user preferences are stable and the relationship between users will not change frequently. Hence, we can update the user relationship graph less frequently than every communication round or update the subgraph instead of the complete graph to improve efficiency. Based on the user relationship graph, we now utilize the full-batch GNN to learn user-specific item embeddings. To further improve the scalability, we can adopt the widely used neighbor sampling strategy\cite{chen2018fastgcn,chiang2019cluster} and only propagate the subgraph to reduce the computation complexity.

\subsubsection{Dynamic and cold-start recommendation.}
Our \baby is a general framework that can be easily extended to various recommendation scenarios. For example, in the sequential recommendation~\cite{kang2018self,chen2022intent} or the session-based recommendation~\cite{wu2019session,pan2022collaborative}, the user interactions are generated dynamically according to timing. On the client side, we can employ a Transformer architecture to capture the sequential properties of data. On the server side, the user relationship graph can be updated adaptively to record the dynamic user preferences. In addition, our model has the capability of handling the cold-start problem~\cite{du2022metakg}. For a new user with limited interactions, our method can discover neighbor users with similar preferences and learn user-specific item embedding to help the new user make recommendations. Compared with other FedRec models, which adopt the common item embedding to recommend, our method can select the most related users to foster the preference depiction of new users.

\begin{table}[!t]
\renewcommand\arraystretch{1.}
\setlength\tabcolsep{0.8pt}
\setlength{\abovecaptionskip}{-6pt}
\centering
% \normalsize
\begin{tabular}{p{75pt}p{35pt}<{\centering}p{35pt}<{\centering}p{60pt}<{\centering}p{35pt}<{\centering}}
\hline
\pmb{Dataset} & \# Users & \# Items & \# Interactions & Sparsity \\
\hline
MovieLens-100K & 943 & 1682 & 100,000 & 93.70\% \\
MovieLens-1M & 6,040 & 3,706 & 1,000,209 & 95.53\% \\
Lastfm-2K & 1,600 & 12,454 & 185,650 & 99.07\% \\
HetRec2011 & 2,113 & 10,109 & 855,598 & 95.99\% \\
Douban & 2,509 & 39,576 & 893,575 & 99.10\% \\
\hline
\end{tabular}
\caption{Dataset statistics.}
\label{datasets}
\end{table}
\section{Experiment}\label{experiment}
In this section, we conduct experiments to analyze the proposed method, aiming to answer below questions:
\begin{itemize}[leftmargin=*]
\item \textbf{\textit{Q1}}: Does \baby outperform the state-of-the-art federated and centralized recommendation models?

\item \textbf{\textit{Q2}}: 
How does our proposed graph learning-based federated recommendation method work?

\item \textbf{\textit{Q3}}: Can the proposed graph-guided aggregation mechanism benefit other FedRec models?

\item \textbf{\textit{Q4}}: How do the key hyper-parameters of \baby impact the performance?

\item \textbf{\textit{Q5}}: Is \baby robust when integrating local differential privacy technique?
\end{itemize}

\subsection{Datasets and Evaluation Protocols}
\paragraph{\textbf{Datasets.}}
We verify the proposed \baby on five recommendation benchmark datasets: MovieLens-100K, MovieLens-1M \cite{harper2015movielens}, Lastfm-2K \cite{Cantador:RecSys2011}, HetRec2011~\cite{Cantador:RecSys2011} and Douban~\cite{hu2014your}. Particularly, two MovieLens datasets are collected from the MovieLens website, which records users' ratings about movies and each user has no less than 20 ratings. Lastfm-2K is a music dataset, where each user retains the listened artists list and listening count. We remove the users with less than 5 interactions from Lastfm-2K. HetRec2011 is an extension of MovieLens-10M, which links the movies with corresponding web pages at Internet Movie Database (IMDb) and Rotten Tomatoes movie review systems. Douban is another user-movie interaction dataset. Detailed statistics of the five datasets are shown in Table \ref{datasets}.

\paragraph{\textbf{Evaluation protocols.}} 
For a fair comparison, we follow the prevalent leave-one-out evaluation setting \cite{he2017neural} and evaluate the performance with Hit Ratio (HR) and Normalized Discounted Cumulative Gain (NDCG) \cite{he2015trirank} metrics. The results are shown in the unit of 1e-2. More details can be found in \textbf{Appendix~\ref{evaluation_protocols}}.

\begin{table*}[!t]
\setlength\tabcolsep{0.6pt}
\centering
% \scriptsize
\begin{tabular}{p{32pt}p{60pt}p{40pt}<{\centering}p{40pt}<{\centering}p{40pt}<{\centering}p{40pt}<{\centering}p{40pt}<{\centering}p{40pt}<{\centering}p{40pt}<{\centering}p{40pt}<{\centering}p{40pt}<{\centering}p{40pt}<{\centering}}
\hline 
& \multirow{2}{*}{\textbf{Method}} & \multicolumn{2}{c}{\textbf{MovieLens-100K}} &
\multicolumn{2}{c}{\textbf{MovieLens-1M}} &
\multicolumn{2}{c}{\textbf{Lastfm-2K}} &
\multicolumn{2}{c}{\textbf{HetRec2011}} &
\multicolumn{2}{c}{\textbf{Douban}} \\
& & HR@10 & NDCG@10 & HR@10 & NDCG@10 & HR@10 & NDCG@10 & HR@10 & NDCG@10 & HR@10 & NDCG@10 \\
\hline
\multirow{3}{*}{\textbf{CenRec}} & \textbf{MF} & 64.48 & 38.61 & 68.69 & 41.45 & 83.13 & 71.78 & 66.07 & 41.21 & 87.17 & 61.75 \\
& \textbf{NCF} & 64.21 & 37.13 & 64.02 & 38.16 & 82.57 & 68.26 & 64.74 & 39.55 & 87.49 & \underline{62.51} \\
& \textbf{SGL} & 64.90 & 40.02 & 62.60 & 34.13 & 82.37 & 68.59 & 65.12 & 40.18 & -- & -- \\
\hline
\multirow{7}{*}{\textbf{FedRec}} & \textbf{FedMF} & 66.17 & 38.73 & 67.91 & 40.81 & 81.63 & 68.18 & 64.69 & 40.29 & 87.17 & 61.00 \\
& \textbf{FedNCF} & 60.66 & 33.93 & 60.38 & 34.13 & 81.44 & 61.95 & 60.86 & 36.27 & 86.01 & 59.94 \\
& \textbf{FedRecon} & 65.22 & 38.49 & 62.78 & 36.82 & 82.06 & 67.37 & 61.57 & 34.20 & \underline{87.52} & 60.38 \\
& \textbf{MetaMF} & 66.21 & 41.02 & 44.98 & 26.31 & 81.04 & 64.13 & 54.52 & 32.36 & 82.58 & 55.44 \\
& \textbf{PFedRec} & \underline{71.37} & \underline{42.59} & \textbf{73.03} & \underline{44.49} & \underline{82.38} & \underline{73.19} & \underline{67.20} & \underline{42.70} & 87.40 & 61.90 \\
& \textbf{FedLightGCN} & 24.53 & 12.78 & 37.53 & 15.01 & 43.75 & 15.17 & 22.65 & 7.96 & 35.66 & 12.33 \\
& \textbf{FedPerGNN} & 11.52 & 5.08 & 9.31 & 4.09 & 10.56 & 4.25 & -- & -- & -- & -- \\
\hline
\multirow{2}{*}{\textbf{Ours}} & \textbf{\baby} & \textbf{72.85*} & 43.77* & 72.17 & 43.61 & \textbf{83.44*} & 74.11* & 69.41* & \textbf{43.34*} & \textbf{88.04*} & 63.87* \\
& \textbf{Light\_\baby} & 72.00* & \textbf{43.92*} & 72.95 & \textbf{45.48*} & \textbf{83.44*} & \textbf{74.33*} & \textbf{69.47*} & 43.21* & \textbf{88.04*} & \textbf{64.00*} \\
\hline
\hline
\multicolumn{2}{c}{\textbf{Improvement}} & $\uparrow$ 2.07\% & $\uparrow$ 3.12\% & -- & $\uparrow$ 2.23\% & $\uparrow$ 1.29\% & $\uparrow$ 1.56\% & $\uparrow$ 3.38\% & $\uparrow$ 1.50\% & $\uparrow$ 0.59\% & $\uparrow$ 2.38\% \\
% \hline
\hline
\end{tabular}
%}
\caption{Performance comparison on five datasets. The best results are bold and the best baseline results are underlined. ``CenRec'' and ``FedRec'' represent centralized and federated settings, respectively. FedPerGNN fails to run on HetRec2011 and Douban due to the unacceptable memory allocation (denoted as "--"). ``\textbf{{\Large *}}'' and ``\textbf{Improvement''} indicate the statistically significant improvement (i.e., two-sided t-test with $p<0.05$) and the performance improvement over the best baseline, respectively.}
\label{performance comparison}
\end{table*}
\subsection{Baselines and Implementation Details}
\paragraph{\textbf{Baselines.}}
We compare our method with two branches of baselines, including centralized and federated recommendation models. All the methods conduct recommendations only based on the user-item interaction without other auxiliary information, which is the most fundamental setting. Details about baselines are summarized in \textbf{Appendix~\ref{baselines}}.

\paragraph{\textbf{Implementation details.}}
We implement the methods based on Pytorch framework and the hyperparameter configuration is summarized in \textbf{Appendix~\ref{implementation_details}}. In addition, we develop a lightweight variant of our method, named Light\_\baby. As discussed in subsection~\ref{discussions}, the Light\_\baby can improve the efficiency by reducing the frequency of user relationship graph updates. 

\subsection{Overall Performance (Q1)}
Table \ref{performance comparison} shows the performance of HR@10 and NDCG@10 on five datasets. Next, we summarize the experimental results and discuss several observations.

\textbf{(1) \baby achieves better performance than centralized methods in all settings}. The largest performance increase of HR@10 and NDCG@10 emerges on MovieLens-100K, \ie 13.46\% and 17.88\%, respectively. In the centralized setting, all users share the same item embedding and score function and only keep user embedding for personalization capture. In comparison, our method maintains user embedding and score function as private components to learn user characteristics. Besides, we introduce a graph structure to mine the correlations between clients, which enhances user preference learning and provides better recommendations.

\textbf{(2) Our method outperforms federated recommendation baselines and achieves state-of-the-art results on almost all datasets}. In FedRec, serving all clients with a unified item embedding ignores distinct user preferences, which hinders user personalization capture. PFedRec learns personalized item embedding by finetuning with local data and achieves the second-best performance, which supports our claim that replacing indiscriminate item embedding with user-specific item embedding can improve the recommendation performance. Compared with PFedRec, our method learns user-specific item embeddings for each user based on the adaptive user relationship graph, which absorbs beneficial information from users with similar preferences and achieves better performance. Besides, the lightweight variant Light\_\baby can achieve comparable and even better performance than \baby, which attains a good balance between model efficiency and efficacy.

\begin{table*}[!t]
\setlength\tabcolsep{0.6pt}
\centering
\setlength{\abovecaptionskip}{-3pt}
% \scriptsize
\begin{tabular}{p{105pt}p{38pt}<{\centering}p{40pt}<{\centering}p{38pt}<{\centering}p{40pt}<{\centering}p{38pt}<{\centering}p{40pt}<{\centering}p{38pt}<{\centering}p{40pt}<{\centering}p{38pt}<{\centering}p{38pt}<{\centering}}
\hline 
\multirow{2}{*}{\textbf{Method}} & \multicolumn{2}{c}{\textbf{MovieLens-100K}} &
\multicolumn{2}{c}{\textbf{MovieLens-1M}} &
\multicolumn{2}{c}{\textbf{Lastfm-2K}} &
\multicolumn{2}{c}{\textbf{HetRec2011}} &
\multicolumn{2}{c}{\textbf{Douban}} \\
& HR@10 & NDCG@10 & HR@10 & NDCG@10 & HR@10 & NDCG@10 & HR@10 & NDCG@10 & HR@10 & NDCG@10 \\
\hline
\textbf{FedNCF} & 60.66 & 33.93 & 60.38 & 34.13 & 81.44 & 61.95 & 60.86 & 36.27 & 86.01 & 59.94 \\
\textbf{FedNCF w/ PSF} & 66.38 & 38.85 & 67.14 & 40.22 & 81.81 & 66.75 & 63.51 & 37.87 & 86.97 & 62.66 \\
\textbf{\baby w/ PSF and Init} & 68.68 & 41.12 & 68.26 & 41.10 & 81.83 & 71.75 & 64.32 & 40.61 & 87.13 & 62.92 \\
\textbf{\baby w/ PSF and Reg} & 71.05 & 42.56 & 67.79 & 42.98 & 82.88 & 73.31 & 66.82 & 38.03 & 87.52 & 63.23 \\
\textbf{\baby} & \textbf{72.85} & \textbf{43.77} & \textbf{72.17} & \textbf{43.61} & \textbf{83.44} & \textbf{74.11} & \textbf{69.41} & \textbf{43.34} & \textbf{88.04} & \textbf{63.87} \\
\hline
\end{tabular}
\caption{Ablation study results. ``\baby--Init'' denotes the model without initializing with popular item embedding and ``\baby--Reg'' denotes the model without regularizing with user-specific item embedding.}
\label{ablation}
\end{table*}
\begin{table*}[!t]
\setlength\tabcolsep{0.7pt}
\centering
% \scriptsize
\begin{tabular}{p{55pt}p{40pt}<{\centering}p{40pt}<{\centering}p{40pt}<{\centering}p{40pt}<{\centering}p{40pt}<{\centering}p{40pt}<{\centering}p{40pt}<{\centering}p{40pt}<{\centering}p{40pt}<{\centering}p{40pt}<{\centering}}
\hline 
\multirow{2}{*}{\textbf{Method}} & \multicolumn{2}{c}{\textbf{MovieLens-100K}} &
\multicolumn{2}{c}{\textbf{MovieLens-1M}} &
\multicolumn{2}{c}{\textbf{Lastfm-2K}} &
\multicolumn{2}{c}{\textbf{HetRec2011}} &
\multicolumn{2}{c}{\textbf{Douban}} \\
& HR@10 & NDCG@10 & HR@10 & NDCG@10 & HR@10 & NDCG@10 & HR@10 & NDCG@10 & HR@10 & NDCG@10 \\
\hline
\textbf{FedMF} & 66.17 & 38.73 & 67.91 & 40.81 & 81.63 & 68.18 & 64.69 & 40.29 & 87.17 & 61.00 \\
\textbf{w/ GraphAgg} & 71.79 & 44.20 & 72.15 & 43.69 & 81.88 & 72.01 & 68.81 & 42.00 & 87.33 & 62.05 \\
\textbf{Improvement} & $\uparrow$ 8.49\% & $\uparrow$ 14.12\% & $\uparrow$ 6.24\% & $\uparrow$ 7.06\% & $\uparrow$ 0.31\% & $\uparrow$ 5.62\% & $\uparrow$ 6.37\% & $\uparrow$ 4.24\% & $\uparrow$ 0.18\% & $\uparrow$ 1.72\% \\
\hline
\textbf{FedRecon} & 65.22 & 38.49 & 62.78 & 36.82 & 82.06 & 67.37 & 61.57 & 34.20 & 87.52 & 60.38 \\
\textbf{w/ GraphAgg} & 70.78 & 41.10 & 69.03 & 40.15 & 82.97 & 73.83 & 62.94 & 35.99 & 87.80 & 61.12 \\
\textbf{Improvement} & $\uparrow$ 8.52\% & $\uparrow$ 6.78\% & $\uparrow$ 9.96\% & $\uparrow$ 9.04\% & $\uparrow$ 1.11\% & $\uparrow$ 9.59\% & $\uparrow$ 2.23\% & $\uparrow$ 5.23\% & $\uparrow$ 0.32\% & $\uparrow$ 1.23\% \\
\hline
\textbf{PFedRec} & 71.37 & 42.59 & 73.03 & 44.49 & 82.38 & 73.19 & 67.20 & 42.70 & 87.40 & 61.90 \\
\textbf{w/ GraphAgg} & 72.38 & 43.75 & 73.50 & 44.53 & 82.63 & 73.11 & 70.00 & 42.76 & 87.56 & 62.00 \\
\textbf{Improvement} & $\uparrow$ 1.42\% & $\uparrow$ 2.72\% & $\uparrow$ 0.64\% & $\uparrow$ 0.09\% & $\uparrow$ 0.30\% & -- & $\uparrow$ 4.17\% & $\uparrow$ 0.14\% & $\uparrow$ 0.18\% & 0.16\% \\
\hline
\end{tabular}
%}
\caption{Performance improvement for integrating our graph-guided aggregation mechanism (denoted as GraphAgg) to baseline algorithms. ``Improvement'' denotes the performance gain of the baselines by incorporating GraphAgg.}
\label{compatibility}
\end{table*}
\textbf{(3) Our graph-guided aggregation mechanism demonstrates significant performance advantages over two federated GNN recommendation models}. FedLightGCN employs a GCN on each client as the representation learning model. The local sub-graph contains the user node and the item nodes the user has interacted with, and the neighborhood of each item node is the same. As a result, the item representations obtained through neighbor aggregation lack discriminability, which is not conducive to recommendation prediction. FedPerGNN performs poorly under the implicit feedback recommendation setting, which samples negative items during model training. FedPerGNN finds the high-order neighbors by matching the user's interactions and the negative items mislead the discovery of actual neighbors, which brings an adverse impact on model performance. The negative effect is even more severe in the leave-one-out setting, where each user has more training samples and samples more negative samples.

\paragraph{\textbf{Convergence analysis.}} 
We compare the convergence of our method and baselines, and there are two main conclusions: \textbf{\textit{First}}, on the two MovieLens datasets, our method shows a similar convergence trend to FedNCF due to the similar backbone architecture in the first half of the training process, and outperforms all baselines in the second half. \textbf{\textit{Second}}, our method converges quickly on the Lastfm-2K and Douban datasets. The model convergence comparison and more details are summarized in \textbf{Appendix~\ref{convergence_comparison}}.

\subsection{Ablation Study (Q2)}\label{exp_diff_graph}
\paragraph{\textbf{Model component analysis.}} 
We decouple our GPFedRec into the basic federated learning scheme, incorporating designed components. FedNCF~\cite{perifanis2022federated} serves as the backbone, implementing NCF under the federated learning scheme. Besides, we introduce the personalized score function and graph-guided aggregation mechanism (global item embedding initialization and user-specific item embedding regularization). To evaluate their effectiveness, we compare the performance of FedNCF, FedNCF with personalized score function (FedNCF w/ PSF), FedNCF with personalized score function and global item embedding initialization (FedNCF w/ PSF and Init), FedNCF with personalized score function and user-specific item embedding regularization (FedNCF w/ PSF and Reg), and GPFedRec (FedNCF with personalized score function and graph-guided aggregation).

From the results in Table \ref{ablation}, we can conclude that (1) adding the personalized score function to FedNCF improves performance. (2) integrating global item embedding initialization or user-specific item embedding regularization further enhances model performance. (3) Combining personalized score function and graph-guided aggregation mechanism achieves the best performance. The shared item embedding depicts the globally popular preference and the user-specific item embedding maintains the personalized preferences of users with similar tastes. The two kinds of information cooperate with each other to help the client models absorb common characteristics while retaining personalized descriptions of different clients, which jointly contribute to the model performance.

\paragraph{\textbf{Effect of different graph construction methods.}} 
We conduct experiments to evaluate the effect of different user relationship user construction methods, \ie random graph, graph built with interactions and ours built with item embeddings. Particularly, for the random graph construction, we adopt four commonly used random graph models, including Barab\'{a}si-Albert (BA)~\cite{albert2002statistical}, Watts-Strogatz (WS)~\cite{watts1998collective}, Erd\H{o}s-R\'{e}nyi (ER)~\cite{gilbert1959random} and Regular graph. For the graph built with user historical interactions, we utilize four metrics for similarity calculation, including Cosine, Euclidean, Jaccard, and Pearson. To make a thorough verification, we set different densities of user connections during graph construction. The brief results summary is shown in Figure \ref{diff_graphs} and more details are summarized in \textbf{Appendix~\ref{graph_construction_effect}}. 

In summary, we have three conclusions: \textbf{\textit{First}}, the model whose graph is generated randomly always gets poor performance. \textbf{\textit{Second}}, the model whose graph is built with user historical interactions can achieve new state-of-the-art performance with Euclidean similarity metric under 80\% connection density. \textbf{\textit{Third}}, our method consistently performs better than random graphs while achieving comparable results to graphs built with user historical interactions but in a privacy-preserving way.

\subsection{Compatibility Study (Q3)}
We verify the compatibility of the proposed graph-guided aggregation mechanism by integrating it into other FedRec models. Particularly, we take FedMF, FedRecon and PFedRec as examples and replace their indiscriminate item embedding aggregation with our mechanism. As shown in Table \ref{compatibility}, all models are significantly improved by introducing our proposed mechanism in almost all cases, which emphasizes the necessity of incorporating user-specific preferences into client models. Moreover, our mechanism does not introduce any additional parameters, which shows outstanding compatibility and great potential to enhance FedRec models.

\subsection{Hyper-parameter Analysis (Q4)}
We conduct experiments to analyze the impact of key hyper-parameters of our method on recommendation performance. 

\paragraph{\textbf{\textit{Threshold of neighborhood selection}}}
In each round, the server collects item embeddings from clients and constructs the user relationship graph by calculating user similarities. 
% We keep the links whose weights are larger than a threshold, \ie the mean of all link weights multiplied by a factor. The smaller the factor, the more links are preserved, and vice versa. 
Particularly, we fix all the other parameters and set the threshold from 0 to 2 with an interval of 0.5.
\begin{figure}[!t]
\setlength{\belowcaptionskip}{-15pt}
\setlength{\abovecaptionskip}{-0.5pt}
\centering
\includegraphics[width=0.48\textwidth,height=0.2\textwidth]{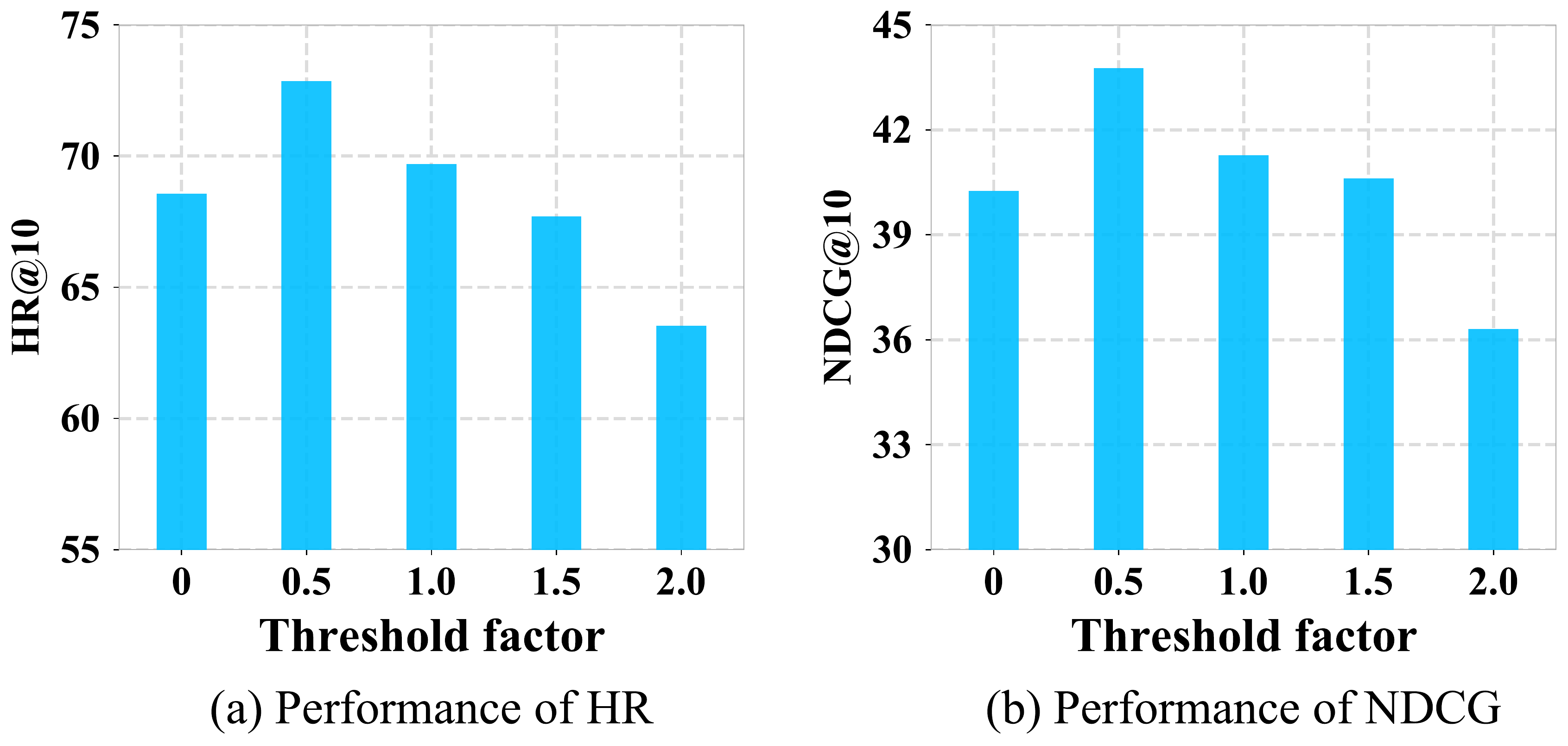}
\caption{Effect of the threshold of neighborhood selection. We show the results of both metrics on MovieLens-100K.}
\label{threshold}
\end{figure}
As shown in Figure \ref{threshold}, we can see that,

(1) As the threshold increases, performance first gets better and then decreases, and the best result is achieved when the factor is 0.5.

(2) When the threshold is 0, it means that every user has links to all other users, and the user relationship graph is fully connected. Then, the common item embedding learned by the server is the average of item embeddings uploaded by all users. Therefore, all users are trained with the same regularization, which constrains the locally updated item embedding not to be too far from common preference. Clearly, this indiscriminate constraint does not help users capture personalized preferences much. 

(3) When the threshold increases, \eg 2.0, each user has fewer neighbors in the user relationship graph. As a result, the user-specific item embedding learned by the server for each user is biased and cannot well characterize the personalized user preferences, which leads to a decrease in model performance.

\begin{figure}[!t]
\setlength{\belowcaptionskip}{-15pt}
\setlength{\abovecaptionskip}{-0.5pt}
\centering
\includegraphics[width=0.48\textwidth,height=0.2\textwidth]{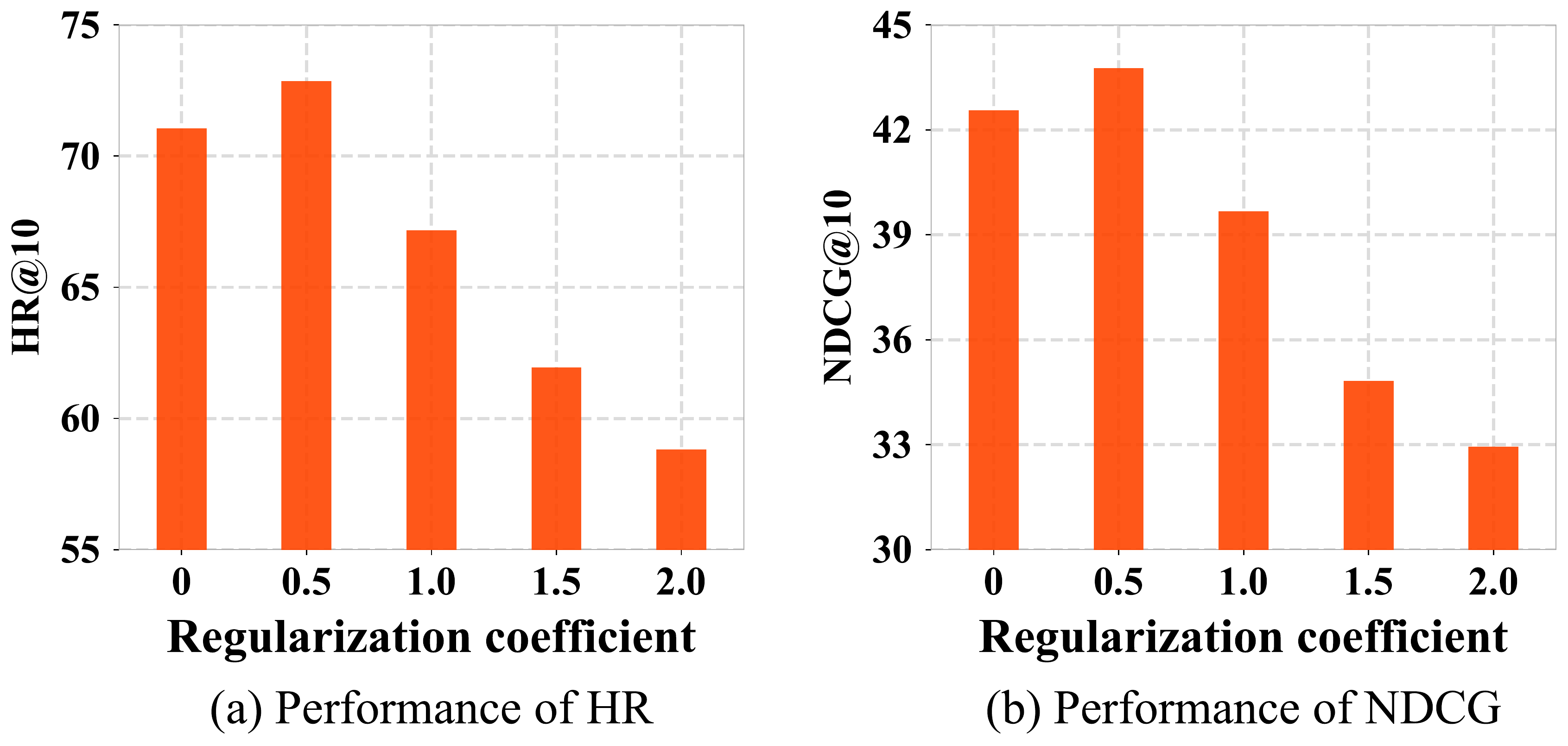}
\caption{Effect of coefficient of the regularization term. We show the results of both metrics on MovieLens-100K.}
\label{reg}
\end{figure}

\paragraph{\textbf{Coefficient of the regularization term}}
In our method, we set a regularization term for the local model training, which offers the client user-specific item embedding from users with similar preferences. 
% We explore the effect of the regularization coefficient value on the model performance. 
Specifically, we fix all the other parameters and set the coefficient from 0 to 2 with an interval of 0.5. From Figure \ref{reg} we can conclude that,

(1) The model performance first increases and then decreases with the raising of the regularization coefficient and the best result appears when the coefficient is 0.5.

(2) When the coefficient is 0, we can see that the performance is also better than almost all baselines. In our method, the global item embedding is calculated by user-specific embeddings obtained with the graph-guided aggregation mechanism. Compared with the indiscriminate aggregation of baseline models, it gives higher weight to popular user preferences which can retain beneficial information for recommendation.

(3) Large coefficients will degrade model performance. The regularization term constrains model refers to users with similar preferences, and the loss function guides the model to capture user personalization based on local data. The too-large coefficient can deviate the user from her own preferences, which in turn interferes with model training.

\begin{figure}[t]
\setlength{\belowcaptionskip}{-6pt}
\setlength{\abovecaptionskip}{-0.5pt}
\centering
\includegraphics[width=0.48\textwidth,height=0.2\textwidth]{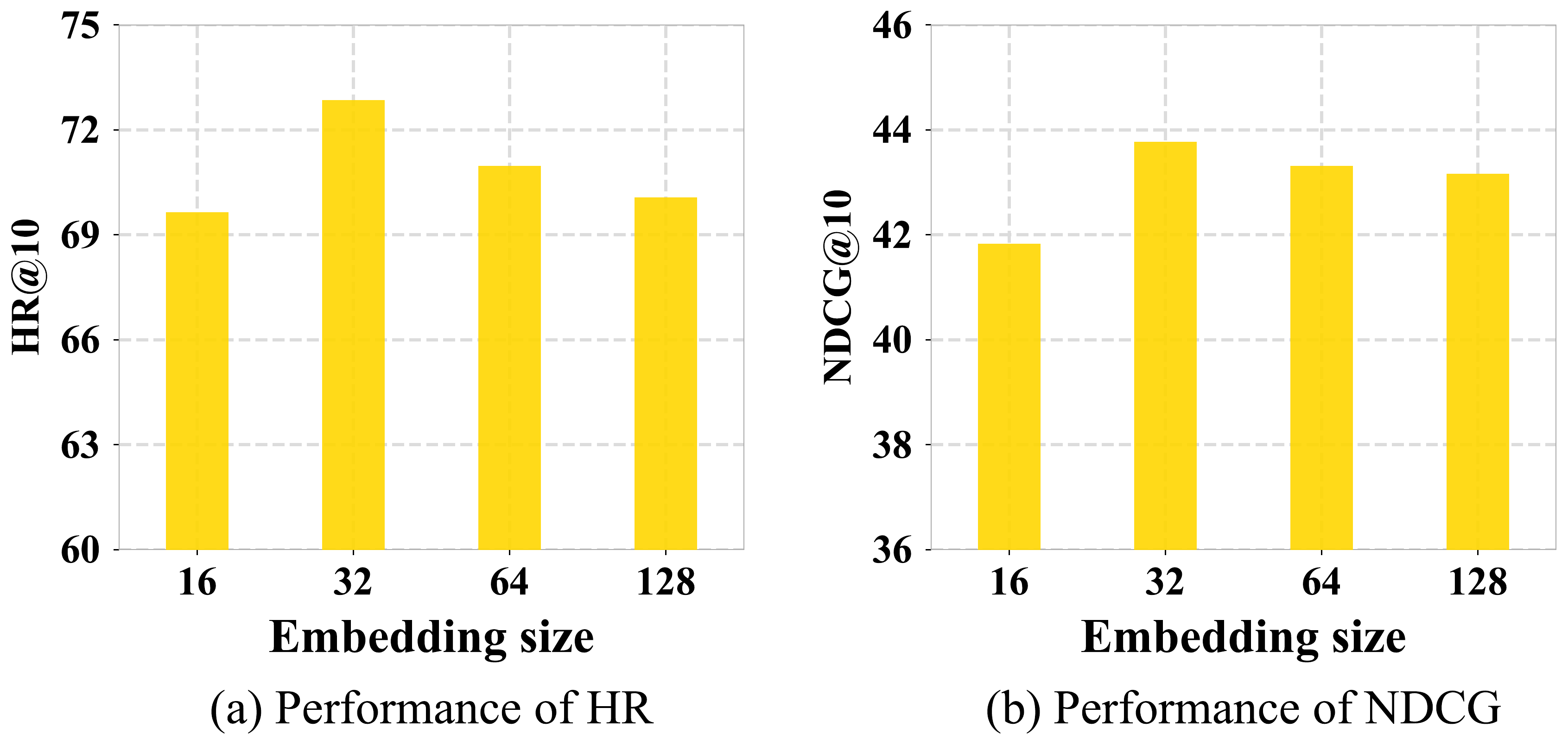}
\caption{Effect of embedding size. We show the results of both metrics on MovieLens-100K.}
\label{emb_size}
\end{figure}

\paragraph{\textbf{Size of embedding}}
Our method employs an MLP as the score function module to predict the user's preference based on the user embedding and item embedding. We fix the MLP architecture and test the effect of different embedding sizes. Particularly, we set the embedding size as 16, 32, 64, and 128, respectively, and the results are summarized in Figure \ref{emb_size}. When the embedding size is 16, the model performance is worse than the others due to limited capacity. As the embedding size grows, the model performance improves accordingly. However, if the dimension is too large, \eg 128, the performance will degrade caused of overfitting.

\subsection{Privacy Protection (Q5)}
\label{sec_exp:privacy}
In this subsection, we evaluate the performance of our privacy protection enhanced \baby with the local differential privacy strategy. Particularly, we set the noise intensity $\delta=[0,0.1,0.2,0.3,0.4,0.5]$ and experimental results are shown in Table \ref{dp}. We can see that the performance declines as the noise intensity $\delta$ grows, while the performance drop is slight if $\delta$ is not too large. Hence, a moderate strength of $\delta$ such as $0.3$ is desirable to achieve a good balance between recommendation accuracy and privacy protection.
\begin{table}[htbp]
\setlength{\abovecaptionskip}{-2mm}
\centering
% \scriptsize
\begin{tabular}{p{50pt}p{20pt}<{\centering}p{20pt}<{\centering}p{20pt}<{\centering}p{20pt}<{\centering}p{20pt}<{\centering}p{20pt}<{\centering}}
\hline
\textbf{Intensity} $\bm{\delta}$ & $0$ & \textbf{$0.1$} & \textbf{$0.2$} & \textbf{$0.3$} & \textbf{$0.4$} & \textbf{$0.5$}  \\
\hline
\textbf{HR@10} & 72.85 & 71.89 & 71.32 & 70.41 & 69.99 & 69.35 \\
\textbf{NDCG@10} & 43.77 & 42.58 & 41.79 & 41.78 & 40.68 & 39.89 \\
\hline
\end{tabular}
%}
\caption{Results of applying local differential privacy technique into our method with various Laplacian noise intensity $\bm{\delta}$.}
\label{dp}
\end{table}

\section{Conclusion}
In this paper, we present a novel graph-guided personalization framework for federated recommendation, named \baby. Our method recovers correlations between users by constructing a user relationship graph on the server. To avoid the potential privacy exposure risk, we build the graph using public item embeddings without collecting private interaction data. We then employ a graph-guided aggregation mechanism to learn many user-specific item embeddings, which enhances user preference modeling. Extensive experiments demonstrate the superior performance gain beyond state-of-the-art baselines. Furthermore, in-depth experiments verify the compatibility of combining our mechanism with other FedRec methods and the robustness of integrating privacy protection techniques into our method, which sheds light on the privacy-preserving federated recommendation deployment in the physical application.

\begin{acks}
Chunxu Zhang and Bo Yang are supported by the National Natural Science Foundation of China under Grant Nos. U22A2098, 62172185, 62206105 and 62202200; the Key Science and Technology Development Plan of Jilin Province under Grant No. 20240302078GX; the National Science and Technology Major Project under Grant No. 2021ZD0112500; the Fundamental Research Funds for the Central Universities, JLU.
\end{acks}

%%
%% The next two lines define the bibliography style to be used, and
%% the bibliography file.
\bibliographystyle{ACM-Reference-Format}
% \balance
\bibliography{sample-base}

%%% -*-BibTeX-*-
%%% Do NOT edit. File created by BibTeX with style
%%% ACM-Reference-Format-Journals [18-Jan-2012].

\begin{thebibliography}{51}

%%% ====================================================================
%%% NOTE TO THE USER: you can override these defaults by providing
%%% customized versions of any of these macros before the \bibliography
%%% command.  Each of them MUST provide its own final punctuation,
%%% except for \shownote{}, \showDOI{}, and \showURL{}.  The latter two
%%% do not use final punctuation, in order to avoid confusing it with
%%% the Web address.
%%%
%%% To suppress output of a particular field, define its macro to expand
%%% to an empty string, or better, \unskip, like this:
%%%
%%% \newcommand{\showDOI}[1]{\unskip}   % LaTeX syntax
%%%
%%% \def \showDOI #1{\unskip}           % plain TeX syntax
%%%
%%% ====================================================================

\ifx \showCODEN    \undefined \def \showCODEN     #1{\unskip}     \fi
\ifx \showDOI      \undefined \def \showDOI       #1{#1}\fi
\ifx \showISBNx    \undefined \def \showISBNx     #1{\unskip}     \fi
\ifx \showISBNxiii \undefined \def \showISBNxiii  #1{\unskip}     \fi
\ifx \showISSN     \undefined \def \showISSN      #1{\unskip}     \fi
\ifx \showLCCN     \undefined \def \showLCCN      #1{\unskip}     \fi
\ifx \shownote     \undefined \def \shownote      #1{#1}          \fi
\ifx \showarticletitle \undefined \def \showarticletitle #1{#1}   \fi
\ifx \showURL      \undefined \def \showURL       {\relax}        \fi
% The following commands are used for tagged output and should be
% invisible to TeX
\providecommand\bibfield[2]{#2}
\providecommand\bibinfo[2]{#2}
\providecommand\natexlab[1]{#1}
\providecommand\showeprint[2][]{arXiv:#2}

\bibitem[Albert and Barab{\'a}si(2002)]%
        {albert2002statistical}
\bibfield{author}{\bibinfo{person}{R{\'e}ka Albert} {and} \bibinfo{person}{Albert-L{\'a}szl{\'o} Barab{\'a}si}.} \bibinfo{year}{2002}\natexlab{}.
\newblock \showarticletitle{Statistical mechanics of complex networks}.
\newblock \bibinfo{journal}{\emph{Reviews of modern physics}} \bibinfo{volume}{74}, \bibinfo{number}{1} (\bibinfo{year}{2002}), \bibinfo{pages}{47}.
\newblock


\bibitem[Ammad-Ud-Din et~al\mbox{.}(2019)]%
        {ammad2019federated}
\bibfield{author}{\bibinfo{person}{Muhammad Ammad-Ud-Din}, \bibinfo{person}{Elena Ivannikova}, \bibinfo{person}{Suleiman~A Khan}, \bibinfo{person}{Were Oyomno}, \bibinfo{person}{Qiang Fu}, \bibinfo{person}{Kuan~Eeik Tan}, {and} \bibinfo{person}{Adrian Flanagan}.} \bibinfo{year}{2019}\natexlab{}.
\newblock \showarticletitle{Federated collaborative filtering for privacy-preserving personalized recommendation system}.
\newblock \bibinfo{journal}{\emph{arXiv preprint arXiv:1901.09888}} (\bibinfo{year}{2019}).
\newblock


\bibitem[Cantador et~al\mbox{.}(2011)]%
        {Cantador:RecSys2011}
\bibfield{author}{\bibinfo{person}{Iv\'{a}n Cantador}, \bibinfo{person}{Peter Brusilovsky}, {and} \bibinfo{person}{Tsvi Kuflik}.} \bibinfo{year}{2011}\natexlab{}.
\newblock \showarticletitle{2nd Workshop on Information Heterogeneity and Fusion in Recommender Systems (HetRec 2011)}. In \bibinfo{booktitle}{\emph{Proceedings of the 5th ACM conference on Recommender systems}} (Chicago, IL, USA) \emph{(\bibinfo{series}{RecSys 2011})}. \bibinfo{publisher}{ACM}, \bibinfo{address}{New York, NY, USA}.
\newblock


\bibitem[Chai et~al\mbox{.}(2020)]%
        {chai2020secure}
\bibfield{author}{\bibinfo{person}{Di Chai}, \bibinfo{person}{Leye Wang}, \bibinfo{person}{Kai Chen}, {and} \bibinfo{person}{Qiang Yang}.} \bibinfo{year}{2020}\natexlab{}.
\newblock \showarticletitle{Secure federated matrix factorization}.
\newblock \bibinfo{journal}{\emph{IEEE Intelligent Systems}} \bibinfo{volume}{36}, \bibinfo{number}{5} (\bibinfo{year}{2020}), \bibinfo{pages}{11--20}.
\newblock


\bibitem[Chen et~al\mbox{.}(2018)]%
        {chen2018fastgcn}
\bibfield{author}{\bibinfo{person}{Jie Chen}, \bibinfo{person}{Tengfei Ma}, {and} \bibinfo{person}{Cao Xiao}.} \bibinfo{year}{2018}\natexlab{}.
\newblock \showarticletitle{FastGCN: Fast learning with graph convolu-tional networks via importance sampling}. In \bibinfo{booktitle}{\emph{International Conference on Learning Representations}}. International Conference on Learning Representations, ICLR.
\newblock


\bibitem[Chen et~al\mbox{.}(2022)]%
        {chen2022intent}
\bibfield{author}{\bibinfo{person}{Yongjun Chen}, \bibinfo{person}{Zhiwei Liu}, \bibinfo{person}{Jia Li}, \bibinfo{person}{Julian McAuley}, {and} \bibinfo{person}{Caiming Xiong}.} \bibinfo{year}{2022}\natexlab{}.
\newblock \showarticletitle{Intent contrastive learning for sequential recommendation}. In \bibinfo{booktitle}{\emph{Proceedings of the ACM Web Conference 2022}}. \bibinfo{pages}{2172--2182}.
\newblock


\bibitem[Chiang et~al\mbox{.}(2019)]%
        {chiang2019cluster}
\bibfield{author}{\bibinfo{person}{Wei-Lin Chiang}, \bibinfo{person}{Xuanqing Liu}, \bibinfo{person}{Si Si}, \bibinfo{person}{Yang Li}, \bibinfo{person}{Samy Bengio}, {and} \bibinfo{person}{Cho-Jui Hsieh}.} \bibinfo{year}{2019}\natexlab{}.
\newblock \showarticletitle{Cluster-gcn: An efficient algorithm for training deep and large graph convolutional networks}. In \bibinfo{booktitle}{\emph{Proceedings of the 25th ACM SIGKDD international conference on knowledge discovery \& data mining}}. \bibinfo{pages}{257--266}.
\newblock


\bibitem[Choi et~al\mbox{.}(2018)]%
        {choi2018guaranteeing}
\bibfield{author}{\bibinfo{person}{Woo-Seok Choi}, \bibinfo{person}{Matthew Tomei}, \bibinfo{person}{Jose Rodrigo~Sanchez Vicarte}, \bibinfo{person}{Pavan~Kumar Hanumolu}, {and} \bibinfo{person}{Rakesh Kumar}.} \bibinfo{year}{2018}\natexlab{}.
\newblock \showarticletitle{Guaranteeing local differential privacy on ultra-low-power systems}. In \bibinfo{booktitle}{\emph{2018 ACM/IEEE 45th Annual International Symposium on Computer Architecture (ISCA)}}. IEEE, \bibinfo{pages}{561--574}.
\newblock


\bibitem[Du et~al\mbox{.}(2022)]%
        {du2022metakg}
\bibfield{author}{\bibinfo{person}{Yuntao Du}, \bibinfo{person}{Xinjun Zhu}, \bibinfo{person}{Lu Chen}, \bibinfo{person}{Ziquan Fang}, {and} \bibinfo{person}{Yunjun Gao}.} \bibinfo{year}{2022}\natexlab{}.
\newblock \showarticletitle{Metakg: Meta-learning on knowledge graph for cold-start recommendation}.
\newblock \bibinfo{journal}{\emph{IEEE Transactions on Knowledge and Data Engineering}} (\bibinfo{year}{2022}).
\newblock


\bibitem[Gao et~al\mbox{.}(2023)]%
        {gao2023survey}
\bibfield{author}{\bibinfo{person}{Chen Gao}, \bibinfo{person}{Yu Zheng}, \bibinfo{person}{Nian Li}, \bibinfo{person}{Yinfeng Li}, \bibinfo{person}{Yingrong Qin}, \bibinfo{person}{Jinghua Piao}, \bibinfo{person}{Yuhan Quan}, \bibinfo{person}{Jianxin Chang}, \bibinfo{person}{Depeng Jin}, \bibinfo{person}{Xiangnan He}, {et~al\mbox{.}}} \bibinfo{year}{2023}\natexlab{}.
\newblock \showarticletitle{A survey of graph neural networks for recommender systems: Challenges, methods, and directions}.
\newblock \bibinfo{journal}{\emph{ACM Transactions on Recommender Systems}} \bibinfo{volume}{1}, \bibinfo{number}{1} (\bibinfo{year}{2023}), \bibinfo{pages}{1--51}.
\newblock


\bibitem[Gilbert(1959)]%
        {gilbert1959random}
\bibfield{author}{\bibinfo{person}{Edgar~N Gilbert}.} \bibinfo{year}{1959}\natexlab{}.
\newblock \showarticletitle{Random graphs}.
\newblock \bibinfo{journal}{\emph{The Annals of Mathematical Statistics}} \bibinfo{volume}{30}, \bibinfo{number}{4} (\bibinfo{year}{1959}), \bibinfo{pages}{1141--1144}.
\newblock


\bibitem[Guo et~al\mbox{.}(2020)]%
        {guo2020survey}
\bibfield{author}{\bibinfo{person}{Qingyu Guo}, \bibinfo{person}{Fuzhen Zhuang}, \bibinfo{person}{Chuan Qin}, \bibinfo{person}{Hengshu Zhu}, \bibinfo{person}{Xing Xie}, \bibinfo{person}{Hui Xiong}, {and} \bibinfo{person}{Qing He}.} \bibinfo{year}{2020}\natexlab{}.
\newblock \showarticletitle{A survey on knowledge graph-based recommender systems}.
\newblock \bibinfo{journal}{\emph{IEEE Transactions on Knowledge and Data Engineering}} \bibinfo{volume}{34}, \bibinfo{number}{8} (\bibinfo{year}{2020}), \bibinfo{pages}{3549--3568}.
\newblock


\bibitem[Harper and Konstan(2015)]%
        {harper2015movielens}
\bibfield{author}{\bibinfo{person}{F~Maxwell Harper} {and} \bibinfo{person}{Joseph~A Konstan}.} \bibinfo{year}{2015}\natexlab{}.
\newblock \showarticletitle{The movielens datasets: History and context}.
\newblock \bibinfo{journal}{\emph{Acm transactions on interactive intelligent systems (tiis)}} \bibinfo{volume}{5}, \bibinfo{number}{4} (\bibinfo{year}{2015}), \bibinfo{pages}{1--19}.
\newblock


\bibitem[He et~al\mbox{.}(2015)]%
        {he2015trirank}
\bibfield{author}{\bibinfo{person}{Xiangnan He}, \bibinfo{person}{Tao Chen}, \bibinfo{person}{Min-Yen Kan}, {and} \bibinfo{person}{Xiao Chen}.} \bibinfo{year}{2015}\natexlab{}.
\newblock \showarticletitle{Trirank: Review-aware explainable recommendation by modeling aspects}. In \bibinfo{booktitle}{\emph{Proceedings of the 24th ACM international on conference on information and knowledge management}}. \bibinfo{pages}{1661--1670}.
\newblock


\bibitem[He et~al\mbox{.}(2020)]%
        {he2020lightgcn}
\bibfield{author}{\bibinfo{person}{Xiangnan He}, \bibinfo{person}{Kuan Deng}, \bibinfo{person}{Xiang Wang}, \bibinfo{person}{Yan Li}, \bibinfo{person}{Yongdong Zhang}, {and} \bibinfo{person}{Meng Wang}.} \bibinfo{year}{2020}\natexlab{}.
\newblock \showarticletitle{Lightgcn: Simplifying and powering graph convolution network for recommendation}. In \bibinfo{booktitle}{\emph{Proceedings of the 43rd International ACM SIGIR conference on research and development in Information Retrieval}}. \bibinfo{pages}{639--648}.
\newblock


\bibitem[He et~al\mbox{.}(2017)]%
        {he2017neural}
\bibfield{author}{\bibinfo{person}{Xiangnan He}, \bibinfo{person}{Lizi Liao}, \bibinfo{person}{Hanwang Zhang}, \bibinfo{person}{Liqiang Nie}, \bibinfo{person}{Xia Hu}, {and} \bibinfo{person}{Tat-Seng Chua}.} \bibinfo{year}{2017}\natexlab{}.
\newblock \showarticletitle{Neural collaborative filtering}. In \bibinfo{booktitle}{\emph{Proceedings of the 26th international conference on world wide web}}. \bibinfo{pages}{173--182}.
\newblock


\bibitem[Hu et~al\mbox{.}(2014)]%
        {hu2014your}
\bibfield{author}{\bibinfo{person}{Longke Hu}, \bibinfo{person}{Aixin Sun}, {and} \bibinfo{person}{Yong Liu}.} \bibinfo{year}{2014}\natexlab{}.
\newblock \showarticletitle{Your neighbors affect your ratings: on geographical neighborhood influence to rating prediction}. In \bibinfo{booktitle}{\emph{Proceedings of the 37th international ACM SIGIR conference on Research \& development in information retrieval}}. \bibinfo{pages}{345--354}.
\newblock


\bibitem[Kang and McAuley(2018)]%
        {kang2018self}
\bibfield{author}{\bibinfo{person}{Wang-Cheng Kang} {and} \bibinfo{person}{Julian McAuley}.} \bibinfo{year}{2018}\natexlab{}.
\newblock \showarticletitle{Self-attentive sequential recommendation}. In \bibinfo{booktitle}{\emph{2018 IEEE international conference on data mining (ICDM)}}. IEEE, \bibinfo{pages}{197--206}.
\newblock


\bibitem[Koren(2008)]%
        {koren2008factorization}
\bibfield{author}{\bibinfo{person}{Yehuda Koren}.} \bibinfo{year}{2008}\natexlab{}.
\newblock \showarticletitle{Factorization meets the neighborhood: a multifaceted collaborative filtering model}. In \bibinfo{booktitle}{\emph{Proceedings of the 14th ACM SIGKDD international conference on Knowledge discovery and data mining}}. \bibinfo{pages}{426--434}.
\newblock


\bibitem[Koren et~al\mbox{.}(2009)]%
        {koren2009matrix}
\bibfield{author}{\bibinfo{person}{Yehuda Koren}, \bibinfo{person}{Robert Bell}, {and} \bibinfo{person}{Chris Volinsky}.} \bibinfo{year}{2009}\natexlab{}.
\newblock \showarticletitle{Matrix factorization techniques for recommender systems}.
\newblock \bibinfo{journal}{\emph{Computer}} \bibinfo{volume}{42}, \bibinfo{number}{8} (\bibinfo{year}{2009}), \bibinfo{pages}{30--37}.
\newblock


\bibitem[Latifi and Jannach(2022)]%
        {latifi2022streaming}
\bibfield{author}{\bibinfo{person}{Sara Latifi} {and} \bibinfo{person}{Dietmar Jannach}.} \bibinfo{year}{2022}\natexlab{}.
\newblock \showarticletitle{Streaming Session-Based Recommendation: When Graph Neural Networks meet the Neighborhood}. In \bibinfo{booktitle}{\emph{Proceedings of the 16th ACM Conference on Recommender Systems}}. \bibinfo{pages}{420--426}.
\newblock


\bibitem[Li et~al\mbox{.}(2023)]%
        {li2023federated}
\bibfield{author}{\bibinfo{person}{Zhiwei Li}, \bibinfo{person}{Guodong Long}, {and} \bibinfo{person}{Tianyi Zhou}.} \bibinfo{year}{2023}\natexlab{}.
\newblock \showarticletitle{Federated recommendation with additive personalization}.
\newblock \bibinfo{journal}{\emph{arXiv preprint arXiv:2301.09109}} (\bibinfo{year}{2023}).
\newblock


\bibitem[Lin et~al\mbox{.}(2020)]%
        {lin2020meta}
\bibfield{author}{\bibinfo{person}{Yujie Lin}, \bibinfo{person}{Pengjie Ren}, \bibinfo{person}{Zhumin Chen}, \bibinfo{person}{Zhaochun Ren}, \bibinfo{person}{Dongxiao Yu}, \bibinfo{person}{Jun Ma}, \bibinfo{person}{Maarten~de Rijke}, {and} \bibinfo{person}{Xiuzhen Cheng}.} \bibinfo{year}{2020}\natexlab{}.
\newblock \showarticletitle{Meta matrix factorization for federated rating predictions}. In \bibinfo{booktitle}{\emph{Proceedings of the 43rd International ACM SIGIR Conference on Research and Development in Information Retrieval}}. \bibinfo{pages}{981--990}.
\newblock


\bibitem[Liu et~al\mbox{.}(2022)]%
        {liu2022federated}
\bibfield{author}{\bibinfo{person}{Zhiwei Liu}, \bibinfo{person}{Liangwei Yang}, \bibinfo{person}{Ziwei Fan}, \bibinfo{person}{Hao Peng}, {and} \bibinfo{person}{Philip~S Yu}.} \bibinfo{year}{2022}\natexlab{}.
\newblock \showarticletitle{Federated social recommendation with graph neural network}.
\newblock \bibinfo{journal}{\emph{ACM Transactions on Intelligent Systems and Technology (TIST)}} \bibinfo{volume}{13}, \bibinfo{number}{4} (\bibinfo{year}{2022}), \bibinfo{pages}{1--24}.
\newblock


\bibitem[McMahan et~al\mbox{.}(2017)]%
        {mcmahan2017communication}
\bibfield{author}{\bibinfo{person}{Brendan McMahan}, \bibinfo{person}{Eider Moore}, \bibinfo{person}{Daniel Ramage}, \bibinfo{person}{Seth Hampson}, {and} \bibinfo{person}{Blaise~Aguera y Arcas}.} \bibinfo{year}{2017}\natexlab{}.
\newblock \showarticletitle{Communication-efficient learning of deep networks from decentralized data}. In \bibinfo{booktitle}{\emph{Artificial Intelligence and Statistics}}. PMLR, \bibinfo{pages}{1273--1282}.
\newblock


\bibitem[Miao et~al\mbox{.}(2023)]%
        {miao2023task}
\bibfield{author}{\bibinfo{person}{Hao Miao}, \bibinfo{person}{Xiaolong Zhong}, \bibinfo{person}{Jiaxin Liu}, \bibinfo{person}{Yan Zhao}, \bibinfo{person}{Xiangyu Zhao}, \bibinfo{person}{Weizhu Qian}, \bibinfo{person}{Kai Zheng}, {and} \bibinfo{person}{Christian~S Jensen}.} \bibinfo{year}{2023}\natexlab{}.
\newblock \showarticletitle{Task Assignment with Efficient Federated Preference Learning in Spatial Crowdsourcing}.
\newblock \bibinfo{journal}{\emph{IEEE Transactions on Knowledge and Data Engineering}} (\bibinfo{year}{2023}).
\newblock


\bibitem[Pan et~al\mbox{.}(2022)]%
        {pan2022collaborative}
\bibfield{author}{\bibinfo{person}{Zhiqiang Pan}, \bibinfo{person}{Fei Cai}, \bibinfo{person}{Wanyu Chen}, \bibinfo{person}{Chonghao Chen}, {and} \bibinfo{person}{Honghui Chen}.} \bibinfo{year}{2022}\natexlab{}.
\newblock \showarticletitle{Collaborative graph learning for session-based recommendation}.
\newblock \bibinfo{journal}{\emph{ACM Transactions on Information Systems (TOIS)}} \bibinfo{volume}{40}, \bibinfo{number}{4} (\bibinfo{year}{2022}), \bibinfo{pages}{1--26}.
\newblock


\bibitem[Pan et~al\mbox{.}(2020)]%
        {pan2020star}
\bibfield{author}{\bibinfo{person}{Zhiqiang Pan}, \bibinfo{person}{Fei Cai}, \bibinfo{person}{Wanyu Chen}, \bibinfo{person}{Honghui Chen}, {and} \bibinfo{person}{Maarten De~Rijke}.} \bibinfo{year}{2020}\natexlab{}.
\newblock \showarticletitle{Star graph neural networks for session-based recommendation}. In \bibinfo{booktitle}{\emph{Proceedings of the 29th ACM international conference on information \& knowledge management}}. \bibinfo{pages}{1195--1204}.
\newblock


\bibitem[Pei et~al\mbox{.}(2024)]%
        {pei2024memory}
\bibfield{author}{\bibinfo{person}{Hongbin Pei}, \bibinfo{person}{Yuheng Xiong}, \bibinfo{person}{Pinghui Wang}, \bibinfo{person}{Jing Tao}, \bibinfo{person}{Jialun Liu}, \bibinfo{person}{Huiqi Deng}, \bibinfo{person}{Jie Ma}, {and} \bibinfo{person}{Xiaohong Guan}.} \bibinfo{year}{2024}\natexlab{}.
\newblock \showarticletitle{Memory Disagreement: A Pseudo-Labeling Measure from Training Dynamics for Semi-supervised Graph Learning}. In \bibinfo{booktitle}{\emph{Proceedings of the ACM on Web Conference 2024}}. \bibinfo{pages}{434--445}.
\newblock


\bibitem[Perifanis and Efraimidis(2022)]%
        {perifanis2022federated}
\bibfield{author}{\bibinfo{person}{Vasileios Perifanis} {and} \bibinfo{person}{Pavlos~S Efraimidis}.} \bibinfo{year}{2022}\natexlab{}.
\newblock \showarticletitle{Federated neural collaborative filtering}.
\newblock \bibinfo{journal}{\emph{Knowledge-Based Systems}}  \bibinfo{volume}{242} (\bibinfo{year}{2022}), \bibinfo{pages}{108441}.
\newblock


\bibitem[Qu et~al\mbox{.}(2023)]%
        {qu2023semi}
\bibfield{author}{\bibinfo{person}{Liang Qu}, \bibinfo{person}{Ningzhi Tang}, \bibinfo{person}{Ruiqi Zheng}, \bibinfo{person}{Quoc Viet~Hung Nguyen}, \bibinfo{person}{Zi Huang}, \bibinfo{person}{Yuhui Shi}, {and} \bibinfo{person}{Hongzhi Yin}.} \bibinfo{year}{2023}\natexlab{}.
\newblock \showarticletitle{Semi-decentralized federated ego graph learning for recommendation}. In \bibinfo{booktitle}{\emph{Proceedings of the ACM Web Conference 2023}}. \bibinfo{pages}{339--348}.
\newblock


\bibitem[Schafer et~al\mbox{.}(2007)]%
        {schafer2007collaborative}
\bibfield{author}{\bibinfo{person}{J~Ben Schafer}, \bibinfo{person}{Dan Frankowski}, \bibinfo{person}{Jon Herlocker}, {and} \bibinfo{person}{Shilad Sen}.} \bibinfo{year}{2007}\natexlab{}.
\newblock \showarticletitle{Collaborative filtering recommender systems}.
\newblock In \bibinfo{booktitle}{\emph{The adaptive web: methods and strategies of web personalization}}. \bibinfo{publisher}{Springer}, \bibinfo{pages}{291--324}.
\newblock


\bibitem[Singhal et~al\mbox{.}(2021)]%
        {singhal2021federated}
\bibfield{author}{\bibinfo{person}{Karan Singhal}, \bibinfo{person}{Hakim Sidahmed}, \bibinfo{person}{Zachary Garrett}, \bibinfo{person}{Shanshan Wu}, \bibinfo{person}{John Rush}, {and} \bibinfo{person}{Sushant Prakash}.} \bibinfo{year}{2021}\natexlab{}.
\newblock \showarticletitle{Federated reconstruction: Partially local federated learning}.
\newblock \bibinfo{journal}{\emph{Advances in Neural Information Processing Systems}}  \bibinfo{volume}{34} (\bibinfo{year}{2021}), \bibinfo{pages}{11220--11232}.
\newblock


\bibitem[Voigt and Von~dem Bussche(2017)]%
        {voigt2017eu}
\bibfield{author}{\bibinfo{person}{Paul Voigt} {and} \bibinfo{person}{Axel Von~dem Bussche}.} \bibinfo{year}{2017}\natexlab{}.
\newblock \showarticletitle{The eu general data protection regulation (gdpr)}.
\newblock \bibinfo{journal}{\emph{A Practical Guide, 1st Ed., Cham: Springer International Publishing}} \bibinfo{volume}{10}, \bibinfo{number}{3152676} (\bibinfo{year}{2017}), \bibinfo{pages}{10--5555}.
\newblock


\bibitem[Wang et~al\mbox{.}(2019)]%
        {wang2019kgat}
\bibfield{author}{\bibinfo{person}{Xiang Wang}, \bibinfo{person}{Xiangnan He}, \bibinfo{person}{Yixin Cao}, \bibinfo{person}{Meng Liu}, {and} \bibinfo{person}{Tat-Seng Chua}.} \bibinfo{year}{2019}\natexlab{}.
\newblock \showarticletitle{Kgat: Knowledge graph attention network for recommendation}. In \bibinfo{booktitle}{\emph{Proceedings of the 25th ACM SIGKDD international conference on knowledge discovery \& data mining}}. \bibinfo{pages}{950--958}.
\newblock


\bibitem[Wang et~al\mbox{.}(2021)]%
        {wang2021learning}
\bibfield{author}{\bibinfo{person}{Xiang Wang}, \bibinfo{person}{Tinglin Huang}, \bibinfo{person}{Dingxian Wang}, \bibinfo{person}{Yancheng Yuan}, \bibinfo{person}{Zhenguang Liu}, \bibinfo{person}{Xiangnan He}, {and} \bibinfo{person}{Tat-Seng Chua}.} \bibinfo{year}{2021}\natexlab{}.
\newblock \showarticletitle{Learning intents behind interactions with knowledge graph for recommendation}. In \bibinfo{booktitle}{\emph{Proceedings of the Web Conference 2021}}. \bibinfo{pages}{878--887}.
\newblock


\bibitem[Watts and Strogatz(1998)]%
        {watts1998collective}
\bibfield{author}{\bibinfo{person}{Duncan~J Watts} {and} \bibinfo{person}{Steven~H Strogatz}.} \bibinfo{year}{1998}\natexlab{}.
\newblock \showarticletitle{Collective dynamics of ‘small-world’networks}.
\newblock \bibinfo{journal}{\emph{nature}} \bibinfo{volume}{393}, \bibinfo{number}{6684} (\bibinfo{year}{1998}), \bibinfo{pages}{440--442}.
\newblock


\bibitem[Wu et~al\mbox{.}(2022b)]%
        {wu2022federated}
\bibfield{author}{\bibinfo{person}{Chuhan Wu}, \bibinfo{person}{Fangzhao Wu}, \bibinfo{person}{Lingjuan Lyu}, \bibinfo{person}{Tao Qi}, \bibinfo{person}{Yongfeng Huang}, {and} \bibinfo{person}{Xing Xie}.} \bibinfo{year}{2022}\natexlab{b}.
\newblock \showarticletitle{A federated graph neural network framework for privacy-preserving personalization}.
\newblock \bibinfo{journal}{\emph{Nature Communications}} \bibinfo{volume}{13}, \bibinfo{number}{1} (\bibinfo{year}{2022}), \bibinfo{pages}{1--10}.
\newblock


\bibitem[Wu et~al\mbox{.}(2021)]%
        {wu2021self}
\bibfield{author}{\bibinfo{person}{Jiancan Wu}, \bibinfo{person}{Xiang Wang}, \bibinfo{person}{Fuli Feng}, \bibinfo{person}{Xiangnan He}, \bibinfo{person}{Liang Chen}, \bibinfo{person}{Jianxun Lian}, {and} \bibinfo{person}{Xing Xie}.} \bibinfo{year}{2021}\natexlab{}.
\newblock \showarticletitle{Self-supervised graph learning for recommendation}. In \bibinfo{booktitle}{\emph{Proceedings of the 44th international ACM SIGIR conference on research and development in information retrieval}}. \bibinfo{pages}{726--735}.
\newblock


\bibitem[Wu et~al\mbox{.}(2020)]%
        {wu2020diffnet++}
\bibfield{author}{\bibinfo{person}{Le Wu}, \bibinfo{person}{Junwei Li}, \bibinfo{person}{Peijie Sun}, \bibinfo{person}{Richang Hong}, \bibinfo{person}{Yong Ge}, {and} \bibinfo{person}{Meng Wang}.} \bibinfo{year}{2020}\natexlab{}.
\newblock \showarticletitle{Diffnet++: A neural influence and interest diffusion network for social recommendation}.
\newblock \bibinfo{journal}{\emph{IEEE Transactions on Knowledge and Data Engineering}} \bibinfo{volume}{34}, \bibinfo{number}{10} (\bibinfo{year}{2020}), \bibinfo{pages}{4753--4766}.
\newblock


\bibitem[Wu et~al\mbox{.}(2019a)]%
        {wu2019neural}
\bibfield{author}{\bibinfo{person}{Le Wu}, \bibinfo{person}{Peijie Sun}, \bibinfo{person}{Yanjie Fu}, \bibinfo{person}{Richang Hong}, \bibinfo{person}{Xiting Wang}, {and} \bibinfo{person}{Meng Wang}.} \bibinfo{year}{2019}\natexlab{a}.
\newblock \showarticletitle{A neural influence diffusion model for social recommendation}. In \bibinfo{booktitle}{\emph{Proceedings of the 42nd international ACM SIGIR conference on research and development in information retrieval}}. \bibinfo{pages}{235--244}.
\newblock


\bibitem[Wu et~al\mbox{.}(2022a)]%
        {wu2022graph}
\bibfield{author}{\bibinfo{person}{Shiwen Wu}, \bibinfo{person}{Fei Sun}, \bibinfo{person}{Wentao Zhang}, \bibinfo{person}{Xu Xie}, {and} \bibinfo{person}{Bin Cui}.} \bibinfo{year}{2022}\natexlab{a}.
\newblock \showarticletitle{Graph neural networks in recommender systems: a survey}.
\newblock \bibinfo{journal}{\emph{Comput. Surveys}} \bibinfo{volume}{55}, \bibinfo{number}{5} (\bibinfo{year}{2022}), \bibinfo{pages}{1--37}.
\newblock


\bibitem[Wu et~al\mbox{.}(2019b)]%
        {wu2019session}
\bibfield{author}{\bibinfo{person}{Shu Wu}, \bibinfo{person}{Yuyuan Tang}, \bibinfo{person}{Yanqiao Zhu}, \bibinfo{person}{Liang Wang}, \bibinfo{person}{Xing Xie}, {and} \bibinfo{person}{Tieniu Tan}.} \bibinfo{year}{2019}\natexlab{b}.
\newblock \showarticletitle{Session-based recommendation with graph neural networks}. In \bibinfo{booktitle}{\emph{Proceedings of the AAAI conference on artificial intelligence}}, Vol.~\bibinfo{volume}{33}. \bibinfo{pages}{346--353}.
\newblock


\bibitem[Yang et~al\mbox{.}(2022)]%
        {yang2022knowledge}
\bibfield{author}{\bibinfo{person}{Yuhao Yang}, \bibinfo{person}{Chao Huang}, \bibinfo{person}{Lianghao Xia}, {and} \bibinfo{person}{Chenliang Li}.} \bibinfo{year}{2022}\natexlab{}.
\newblock \showarticletitle{Knowledge graph contrastive learning for recommendation}. In \bibinfo{booktitle}{\emph{Proceedings of the 45th International ACM SIGIR Conference on Research and Development in Information Retrieval}}. \bibinfo{pages}{1434--1443}.
\newblock


\bibitem[Yin et~al\mbox{.}(2024)]%
        {yin2024device}
\bibfield{author}{\bibinfo{person}{Hongzhi Yin}, \bibinfo{person}{Liang Qu}, \bibinfo{person}{Tong Chen}, \bibinfo{person}{Wei Yuan}, \bibinfo{person}{Ruiqi Zheng}, \bibinfo{person}{Jing Long}, \bibinfo{person}{Xin Xia}, \bibinfo{person}{Yuhui Shi}, {and} \bibinfo{person}{Chengqi Zhang}.} \bibinfo{year}{2024}\natexlab{}.
\newblock \showarticletitle{On-Device Recommender Systems: A Comprehensive Survey}.
\newblock \bibinfo{journal}{\emph{arXiv preprint arXiv:2401.11441}} (\bibinfo{year}{2024}).
\newblock


\bibitem[Yuan et~al\mbox{.}(2023)]%
        {yuan2023interaction}
\bibfield{author}{\bibinfo{person}{Wei Yuan}, \bibinfo{person}{Chaoqun Yang}, \bibinfo{person}{Quoc Viet~Hung Nguyen}, \bibinfo{person}{Lizhen Cui}, \bibinfo{person}{Tieke He}, {and} \bibinfo{person}{Hongzhi Yin}.} \bibinfo{year}{2023}\natexlab{}.
\newblock \showarticletitle{Interaction-level membership inference attack against federated recommender systems}. In \bibinfo{booktitle}{\emph{Proceedings of the ACM Web Conference 2023}}. \bibinfo{pages}{1053--1062}.
\newblock


\bibitem[Zhang et~al\mbox{.}(2023a)]%
        {zhang2023dual}
\bibfield{author}{\bibinfo{person}{Chunxu Zhang}, \bibinfo{person}{Guodong Long}, \bibinfo{person}{Tianyi Zhou}, \bibinfo{person}{Peng Yan}, \bibinfo{person}{Zijian Zhang}, \bibinfo{person}{Chengqi Zhang}, {and} \bibinfo{person}{Bo Yang}.} \bibinfo{year}{2023}\natexlab{a}.
\newblock \showarticletitle{Dual personalization on federated recommendation}. In \bibinfo{booktitle}{\emph{Proceedings of the Thirty-Second International Joint Conference on Artificial Intelligence}}. \bibinfo{pages}{4558--4566}.
\newblock


\bibitem[Zhang et~al\mbox{.}(2024)]%
        {zhang2024federated}
\bibfield{author}{\bibinfo{person}{Chunxu Zhang}, \bibinfo{person}{Guodong Long}, \bibinfo{person}{Tianyi Zhou}, \bibinfo{person}{Zijian Zhang}, \bibinfo{person}{Peng Yan}, {and} \bibinfo{person}{Bo Yang}.} \bibinfo{year}{2024}\natexlab{}.
\newblock \showarticletitle{When Federated Recommendation Meets Cold-Start Problem: Separating Item Attributes and User Interactions}. In \bibinfo{booktitle}{\emph{Proceedings of the ACM on Web Conference 2024}}. \bibinfo{pages}{3632--3642}.
\newblock


\bibitem[Zhang et~al\mbox{.}(2023b)]%
        {zhang2023comprehensive}
\bibfield{author}{\bibinfo{person}{Shijie Zhang}, \bibinfo{person}{Wei Yuan}, {and} \bibinfo{person}{Hongzhi Yin}.} \bibinfo{year}{2023}\natexlab{b}.
\newblock \showarticletitle{Comprehensive privacy analysis on federated recommender system against attribute inference attacks}.
\newblock \bibinfo{journal}{\emph{IEEE Transactions on Knowledge and Data Engineering}} (\bibinfo{year}{2023}).
\newblock


\bibitem[Zhong et~al\mbox{.}(2023)]%
        {zhong2023personalized}
\bibfield{author}{\bibinfo{person}{Xiaolong Zhong}, \bibinfo{person}{Hao Miao}, \bibinfo{person}{Dazhuo Qiu}, \bibinfo{person}{Yan Zhao}, {and} \bibinfo{person}{Kai Zheng}.} \bibinfo{year}{2023}\natexlab{}.
\newblock \showarticletitle{Personalized Location-Preference Learning for Federated Task Assignment in Spatial Crowdsourcing}. In \bibinfo{booktitle}{\emph{Proceedings of the 32nd ACM International Conference on Information and Knowledge Management}}. \bibinfo{pages}{3534--3543}.
\newblock


\bibitem[Zou et~al\mbox{.}(2022)]%
        {zou2022multi}
\bibfield{author}{\bibinfo{person}{Ding Zou}, \bibinfo{person}{Wei Wei}, \bibinfo{person}{Xian-Ling Mao}, \bibinfo{person}{Ziyang Wang}, \bibinfo{person}{Minghui Qiu}, \bibinfo{person}{Feida Zhu}, {and} \bibinfo{person}{Xin Cao}.} \bibinfo{year}{2022}\natexlab{}.
\newblock \showarticletitle{Multi-level cross-view contrastive learning for knowledge-aware recommender system}. In \bibinfo{booktitle}{\emph{Proceedings of the 45th International ACM SIGIR Conference on Research and Development in Information Retrieval}}. \bibinfo{pages}{1358--1368}.
\newblock


\end{thebibliography}

%%
%% If your work has an appendix, this is the place to put it.
\appendix
\section{Evaluation Protocols}\label{evaluation_protocols}
For a fair comparison, we follow the prevalent leave-one-out evaluation setting \cite{he2017neural}. For each user, we take the latest interacted item as the test sample and others for training. Besides, we keep the last reaction of the training set as a validation sample for hyper-parameter selection. To alleviate {the} high computational cost to rank all items for each user during evaluation, we sample 99 items that haven't been interacted with by user and rank the test instance among 100 items, following the common strategy \cite{he2017neural,koren2008factorization}. We evaluate the performance of the ranked list with Hit Ratio (HR) and Normalized Discounted Cumulative Gain (NDCG) \cite{he2015trirank}. To be specific, HR measures whether the test sample is in the top-K list and NDCG assigns higher scores for positions at the top ranks. In this paper, the default list length $K$ is 10.

\section{Baselines}\label{baselines}
We introduce the details of baselines as follows,
\begin{itemize}
    \item \textbf{Matrix Factorization (MF)} \cite{koren2009matrix}: This method is a typical recommendation model. It decomposes the rating matrix into two embeddings in the same latent space to describe user and item characteristics, respectively.
    \item \textbf{Neural Collaborative Filtering (NCF)} \cite{he2017neural}: This method is one of the most representative neural recommendation models. It first learns a user embedding module and an item embedding module, and then employs an MLP to model user-item interaction.
    \item \textbf{Self-supervised Graph Learning (SGL)} \cite{wu2021self}: This method is a self-supervised graph learning enhanced recommendation model. It supplements the traditional supervised recommendation system optimization objective with the auxiliary self-supervised task by constraining the node representation similar under different views.
    \item \textbf{FedMF} \cite{chai2020secure}: It is the federated version of MF, which trains user embedding locally and uploads item gradients to the server for global aggregation.
    \item \textbf{FedNCF} \cite{perifanis2022federated}: It is federated version of NCF. Particularly, it regards user embedding as a private component trained locally and shares item embedding and MLP to perform collaborative training.
    \item \textbf{Federated Reconstruction (FedRecon)} \cite{singhal2021federated}: It is an advanced personalized federated learning framework, and we evaluate it on matrix factorization. Different from FedMF, FedRecon retrains user embedding in each round and computes item gradients based on the retrained user embedding.
    \item \textbf{Meta Matrix Factorization (MetaMF)}~\cite{lin2020meta}: It is a distributed matrix factorization framework where a meta-network is adopted to generate the score function module and private item embedding.
    \item \textbf{Personalized Federated Recommendation (PFedRec)~\cite{zhang2023dual}}: It is a personalized federated recommendation framework where the server first learns a common item embedding for all clients and then each client finetunes the item embedding with local data.
    \item \textbf{Federated LightGCN (FedLightGCN)}: We extend the LightGCN~\cite{he2020lightgcn} to the federated learning framework. Particularly, each client trains the local LightGCN with the first-order interaction subgraph.
    \item \textbf{Federated Graph Neural Network (FedPerGNN)~\cite{wu2022federated}}: It deploys a graph neural network in each client and the user can incorporate high-order user-item information by a graph expansion protocol.
\end{itemize}

\vspace{-3mm}
\section{Convergence Comparison}\label{convergence_comparison}
\begin{figure*}[!t]
{\subfigure{\includegraphics[width=1.\linewidth]{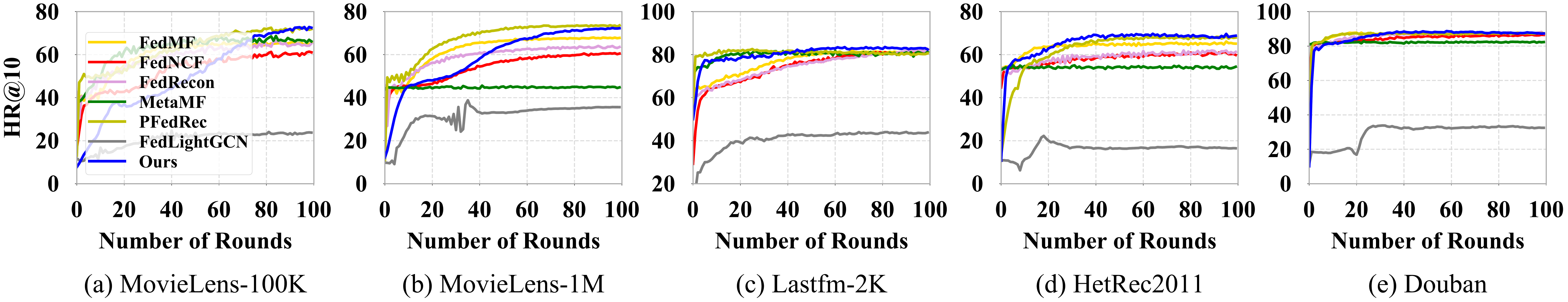}}}

{\subfigure{\includegraphics[width=1.\linewidth]{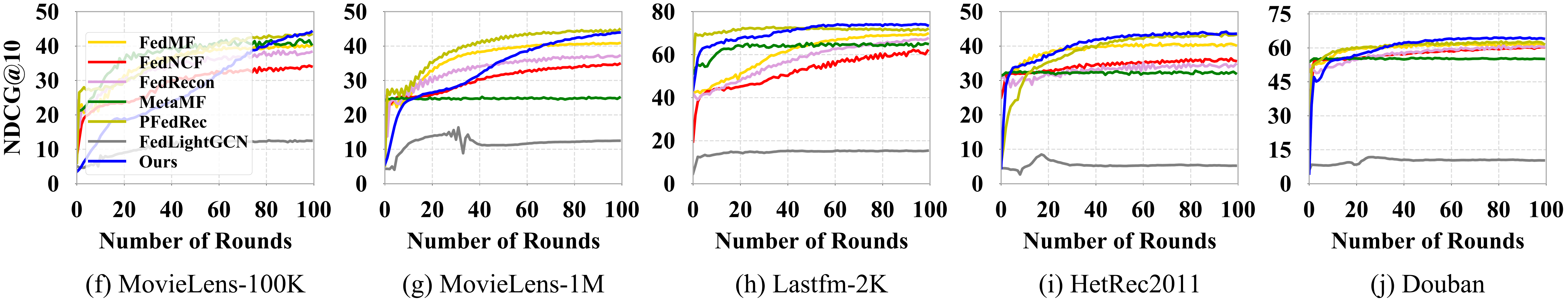}}}
\caption{Model convergence comparison. The horizontal axis is the number of federated optimization rounds, and the vertical axis is the model performance on both metrics.}
\label{convergence}
\end{figure*}

We compare the convergence of our method and baselines and Figure \ref{convergence} illustrates results under two metrics (FedPerGNN and SGL are omitted due to too few iterations). On the two MovieLens datasets, our method shows a similar convergence trend to FedNCF due to the similar backbone architecture in the first half of the training process and outperforms all baselines in the second half. There are more interactions of each user in the two MovieLens datasets. {For those models that capture personalization based only on local data, user preference learning can be achieved quickly in the early stages of training. However, when the model gradually converges with local data, the performance rises slowly. In contrast, our model can leverage user-specific preference information obtained from other users with similar preferences besides local data, which benefits personalization handling and achieves better performance.}

Besides, we can see that our method converges quickly on the Lastfm-2K and Douban datasets. As described in Table 2, the sparsity is as high as 99.07\% for the Lastfm-2k dataset and 99.10\% for the Douban dataset, which means that there are fewer available interaction data for each user to model preference. Our method learns the personalized item embedding by aggregating users with high similarity, which alleviates the difficulty of local personalization modeling and accelerates convergence.

\begin{table*}[!t]
\renewcommand\arraystretch{1.1}
\setlength\tabcolsep{0.6pt}
\centering
\small
\begin{tabular}{p{35pt}|p{28pt}<{\centering}p{28pt}<{\centering}|p{28pt}<{\centering}p{28pt}<{\centering}|p{28pt}<{\centering}p{28pt}<{\centering}|p{28pt}<{\centering}p{28pt}<{\centering}|p{28pt}<{\centering}p{28pt}<{\centering}|p{28pt}<{\centering}p{28pt}<{\centering}|p{28pt}<{\centering}p{28pt}<{\centering}|p{28pt}<{\centering}p{28pt}<{\centering}}
\hline 
\multirow{3}{*}{\textbf{Density}} & \multicolumn{8}{c}{\textbf{Random graph}} & \multicolumn{8}{c}{\textbf{User historical interactions}} \\
\cline{2-17}
& \multicolumn{2}{c}{BA} \vline & \multicolumn{2}{c}{WS} \vline & \multicolumn{2}{c}{ER} \vline & \multicolumn{2}{c}{Regular} \vline & \multicolumn{2}{c}{Cosine} \vline & \multicolumn{2}{c}{Euclidean} \vline & \multicolumn{2}{c}{Jaccard} \vline & \multicolumn{2}{c}{Pearson} \\
\cline{2-17}
& HR & NDCG & HR & NDCG & HR & NDCG & HR & NDCG & HR & NDCG & HR & NDCG & HR & NDCG & HR & NDCG \\
\hline
\textbf{10\%} & \textbf{69.79} & \textbf{40.81} & \textbf{69.52} & 40.76 & 69.20 & 40.53 & \textbf{69.67} & 41.00 & 68.50 & 41.20 & 69.78 & 42.15 & 65.22 & 37.02 & 68.61 & 39.53 \\
\textbf{20\%} & 68.61 & 40.69 & 57.12 & 37.04 & 69.41 & 40.78 & 68.41 & 40.27 & 66.81 & 39.20 & 70.94 & 42.07 & 66.91 & 39.67 & 67.23 & 38.42 \\
\textbf{30\%} & 69.35 & 39.60 & 68.29 & 39.03 & \textbf{70.52} & \textbf{41.71} & 68.63 & \textbf{41.02} & 68.40 & 39.44 & 69.67 & 41.24 & 69.88 & 41.78 & 69.35 & 40.25 \\
\textbf{40\%} & 69.69 & 40.83 & 67.76 & 39.95 & 69.52 & 41.38 & 69.46 & 40.73 & 68.61 & 40.39 & 70.94 & 42.52 & 69.25 & 42.15 & 69.46 & 40.65 \\
\textbf{50\%} & 68.94 & 40.66 & 63.31 & 35.21 & 67.13 & 38.21 & 68.10 & 40.24 & 68.93 & 40.66 & 71.69 & 43.52 & 70.52 & 41.57 & 69.03 & 40.82 \\
\textbf{60\%} & 67.02 & 39.36 & 69.26 & \textbf{41.16} & 70.10 & 41.51 & 69.14 & 39.89 & 70.63 & 40.77 & 70.84 & 41.33 & 68.61 & 39.65 & 71.37 & 43.30 \\
\textbf{70\%} & 67.87 & 40.75 & 65.54 & 38.02 & 69.90 & 39.43 & 68.10 & 40.77 & \textbf{71.79} & \textbf{43.31} & 71.16 & 42.82 & 71.16 & 42.55 & 69.57 & 41.05 \\
\textbf{80\%} & 68.93 & 40.12 & 69.26 & 40.34 & 70.31 & 39.90 & 68.29 & 39.27 & 69.03 & 41.62 & \textbf{73.17} & \textbf{44.73} & \textbf{71.47} & \textbf{43.05} & 70.52 & 43.03 \\
\textbf{90\%} & 63.94 & 36.67 & 69.16 & 40.90 & 69.63 & 41.67 & 69.35 & 39.45 & 69.67 & 42.69 & 69.57 & 41.19 & 70.41 & 42.94 & \textbf{71.90} & \textbf{44.05} \\
\hline
\end{tabular}
\caption{Performance comparison of different user relationship graph construction methods on the MovieLens-100K dataset and the best result of each graph construction method is bold. \textbf{Random graph} denotes simulating the graph with a randomly generated graph. \textbf{User historical interactions} means building the graph by calculating the similarities of user historical interactions. \textbf{Density} represents the user connection density of the graph.}
\label{graph_table}
\end{table*}

\section{Implementation Details}\label{implementation_details}
During training, for each positive instance, we randomly sample 4 negative instances for all methods from the items that haven't been interacted with \cite{he2017neural}. For a fair comparison, we set the embedding size as 32 for all methods, and other model details of the baseline are followed from the original paper. We use a fixed batch size of 256 and search the learning rate in $[0.0001, 0.001, 0.01, 0.1]$ via the validation set performance. We set the total training epochs (for centralized methods) or communication rounds (for federated methods) as 100, which enables all methods to converge. One exception is FedPerGNN, where we follow the experimental setting with the official code in the original paper, whose communication round is set to 3 (In experiments, we found that more rounds of communication did not lead to performance gain). For the score function module in our method, NCF and FedNCF, we employ three hidden layers MLP whose architecture is $32 \rightarrow 16 \rightarrow 8 \rightarrow 1$.

\section{Effect of Different Graph Construction Methods}\label{graph_construction_effect}
We conduct experiments to verify the effect of different user relationship graph construction methods. Particularly, we set the user connection density from 10\% to 90\% with an interval of 10\% to build the graph. Experimental results are summarized in Table \ref{graph_table}. We can see that the model whose graph is generated randomly always gets worse performance than the graph built with user historical interactions. Generally, the performance is better when the user connection graph density is larger.

\end{document}